\documentclass[a4paper,reqno]{amsart}
\usepackage[printedin,extnum]{myamsart}

\usepackage{graphicx}
\input{mydef}

\providecommand{\myhbar}{h}
\providecommand{\orbit}[1]{\mathcal{O}_{#1}}
\providecommand{\uir}[1]{\rho_{#1}}
\providecommand{\oper}[1]{\mathcal{#1}}
\providecommand{\matr}[4]{{\ensuremath{ \left(\!\! \begin{array}{cc}
#1 & #2 \\ #3 & #4
\end{array}\!\!\right) }}}
\providecommand{\twovect}[2]{{\ensuremath{ \left(\!\! \begin{array}{c}
#1 \\ #2 \end{array}\!\!\right) }}}

\providecommand{\circstack}[1]{\stackrel{\scriptscriptstyle\circ}{#1}}

\newcommand{\anti}{\mathcal{A}}

\newcommand{\ub}[3][]{\left\{\!#1\left[#2,#3\right]\!#1\right\}}

\usepackage{pstricks}
\ppnum{\href{http://arXiv.org/abs/quant-ph/0212101}
{arXiv:\texttt{quant-ph/0212101}}
}{LEEDS-PURE-MATH-200\reflectbox{2}-31}
{200\reflectbox{2}}

\providecommand{\fl}{}

\providecommand{\mailto}[1]{\href{mailto:#1}{#1}}

\begin{document}

\title[$p$-Mechanics: An Introduction]{$p$-Mechanics as a Physical
  Theory.\\ An Introduction}
\author[Vladimir V. Kisil]%
{\href{http://maths.leeds.ac.uk/~kisilv/}{Vladimir
    V. Kisil}}

\thanks{On leave from the Odessa University.}

\address{%
School of Mathematics\\
University of Leeds\\
Leeds LS2\,9JT\\
UK}

\email{\mailto{kisilv@maths.leeds.ac.uk}}

\urladdr{\href{http://maths.leeds.ac.uk/~kisilv/}%
{http://maths.leeds.ac.uk/\~{}kisilv/}}

\begin{abstract}
  The paper provides an introduction into \(p\)-mechanics, which is
  a consistent physical theory suitable for a simultaneous description of 
  classical and quantum mechanics. \(p\)-Mechanics
  naturally provides a common ground for several different approaches
  to quantisation (geometric, Weyl, coherent states, Berezin,
  deformation, Moyal, etc.) and has a potential for expansions into
  field and string theories. The backbone of \(p\)-mechanics is solely
  the representation theory of the Heisenberg group.
\end{abstract}

\AMSMSC{81R05}{81R15, 22E27, 22E70, 43A65}

\maketitle

\tableofcontents
\listoffigures


\section{Introduction}
\label{sec:introduction}

This paper describes how classical and quantum mechanics are naturally
united within a construction based on the Heisenberg group
\(\Space{H}{n}\) and the complete set of its unitary representations.
There is a dynamic equation~\eqref{eq:p-equation} on \(\Space{H}{n}\)
which generates both Heisenberg~\eqref{eq:moyal-equation} and
Hamilton~\eqref{eq:Hamilton-equation} equations and corresponding
classical and quantum dynamics. The standard assumption that
observables constitute an algebra, which is discussed
in~\cite{Kisil97a,Kisil02c} and elsewhere, is not necessary for
setting up a valid quantisation scheme.

\emph{The paper outline} is as follows. In the next
Section we recall the representation theory of the Heisenberg group
based on the \emph{orbit method} of Kirillov~\cite{Kirillov99} and
utilising Fock--Segal--Bargmann spaces~\cite{Folland89,Howe80b}. We
emphasise the existence \emph{and} usability of the family of
one-dimensional representations: they play for classical mechanics
exactly the same r\^ole as infinite dimensional representations do for
quantum. In Section~\ref{sec:p-mechanics-statics} we introduce the
concept of observable in \(p\)-mechanics and describe their relations
with quantum and classical observables. These links are provided by
the representations of the Heisenberg group and \emph{wavelet}
transforms. In Section~\ref{sec:p-mechanics-dynamics} we study
\(p\)-mechanical brackets and the associated dynamic equation together
with its classical and quantum representations. In conclusion we
derive the symplectic invariance of dynamics from automorphisms of
\(\Space{H}{n}\). 

The notion of physical states in \(p\)-mechanics is introduced in
subsequent publications~\cite{Brodlie03a,BrodlieKisil03a};
\(p\)-mechanical approach to quantised fields is sketched
in~\cite{Kisil03a} with some further papers to follow.  

\section{The Heisenberg Group and Its Representations}
\label{sec:preliminaries}

We start from the representation theory of the Heisenberg group
\(\Space{H}{n}\) based
on the \emph{orbit method} of Kirillov. Analysis of the unitary dual of
\(\Space{H}{n}\) in Subsection~\ref{sec:struct-topol-unit} suggests
that the family of one-dimensional representations of \(\Space{H}{n}\)
forms the phase space of a classical system. Infinite dimensional
representations in the Fock type space are described in
Subsection~\ref{sec:fock-type-spaces}. 

\subsection{Representations $\Space{H}{n}$ and Method of Orbit}
\label{sec:repr-meth-orbit}

Let \((s,x,y)\), where \(x\), \(y\in \Space{R}{n}\) and \(s\in\Space{R}{}\), be
an element of the Heisenberg group
\(\Space{H}{n}\)~\cite{Folland89,Howe80b}. We assign physical units to
coordinates on \(\Space{H}{n}\). Let \(M\) be a unit of mass,
\(L\)---of length, \(T\)---of time then we adopt the following
\begin{conv}
  \label{conv:units-on-Hn}
  \begin{enumerate}
  \item \label{it:addition} Only physical quantities of the \emph{same
      dimension} can be added or subtracted.
  \item \label{it:functions} Therefore mathematical functions, e.g.
    \(\exp(u)=1+u+u^2/2!+\ldots\) or \(\sin(u)\), can be
    naturally constructed out of a dimensionless number \(u\) only. Thus
    Fourier dual variables, say \(x\) and \(q\), should posses
    reciprocal dimensions because they have to form the expression
    \(e^{\rmi xq}\).
  \item \label{it:a-priori}
    We assign to \(x\) and \(y\) components of \((s,x,y)\) physical
    units \(1/L\) and \(T/(LM)\) respectively.
\end{enumerate}
\end{conv}
  The Convention~\ref{it:a-priori} is the only \emph{a priori}
  assumption which we made about physical dimensions and it will be
  justified \emph{a posteriori} as follows.  From~\ref{it:functions}
  we need dimensionless products \(qx\) and \(py\) in order to get the
  exponent in~\eqref{eq:stone-one}, where \(q\) and \(p\) represent
  the classical coordinates and momenta (in accordance to the main
  observation of \(p\)-mechanics). All other dimensions will be
  assigned strictly in agreement with the
  Convention~\ref{it:addition} and~\ref{it:functions}.

The group law on
\(\Space{H}{n}\) is given as follows:
\begin{equation}
  \label{eq:H-n-group-law}
  (s,x,y)*(s',x',y')=(s+s'+\frac{1}{2} \omega(x,y;x',y'),x+x',y+y'), 
\end{equation} 
where the non-commutativity is solely due to \(\omega\)---the
\emph{symplectic form}~\cite[\S~37]{Arnold91} on the Euclidean space \(\Space{R}{2n}\): 
\begin{equation}
  \label{eq:symplectic-form}
  \omega(x,y;x',y')=xy'-x'y.
\end{equation}
Consequently the parameter \(s\) should be measured in \(T/(L^2
M)\)---the product of units of \(x\) and \(y\).
The Lie algebra \(\algebra{h}^n\) of \(\Space{H}{n}\) is spanned by the
basis \(S\), \(X_j\), \(Y_j\), \(j=1,\ldots,n\), which may be
represented by either left- or right-invariant vector fields on \(\Space{H}{n}\):
\begin{equation}
  S^{l(r)}=\pm\frac{\partial}{\partial s}, \qquad
  X_j^{l(r)}=\pm\frac{\partial}{\partial x_j}-\frac{y_j}{2}\frac{\partial}{\partial s},  \qquad
 Y_j^{l(r)}=\pm\frac{\partial}{\partial y_j}+\frac{x_j}{2}\frac{\partial}{\partial s}.
  \label{eq:h-lie-algebra}
\end{equation}
These fields satisfy the Heisenberg \emph{commutator relations} expressed through the
Kronecker delta \(\delta_{i,j}\) as follows:
\begin{equation}
  \label{eq:heisenberg-comm}
  [X_i^{l(r)},Y_j^{l(r)}]=\delta_{i,j}S^{l(r)} 
\end{equation} and all other commutators (including any between a left
and a right fields) vanishing. Units to measure
\(S^{l(r)}\), \(X_j^{l(r)}\), and \(Y_j^{l(r)}\) are inverse to \(s\),
\(x\), \(y\)---i.e.
\(L^2M/T\), \(L\), and \(LM/T\) respectively---which
are obviously compatible with~\eqref{eq:heisenberg-comm}.

The exponential map \(\exp:
\algebra{h}^n\rightarrow \Space{H}{n}\) respecting
the multiplication~\eqref{eq:H-n-group-law} and
Heisenberg commutators~\eqref{eq:heisenberg-comm} is provided by the formula:
\begin{displaymath}
  \exp: sS+\sum_{j=1}^n (x_jX_j+y_jY_j) \mapsto
  (s,x_1,\ldots,x_n,y_1,\ldots,y_n).
\end{displaymath} The composition of the exponential map with
representations~\eqref{eq:h-lie-algebra} of \(\algebra{h}^n\) by the
left(right)-invariant vector fields produces the right (left) regular
representation \(\lambda_{r(l)}\) of \(\Space{H}{n}\) by right (left)
shifts. Linearised~\cite[\S~7.1]{Kirillov76} to
\(\FSpace{L}{2}(\Space{H}{n})\) they are:
\begin{equation}
  \label{eq:left-right-regular}
  \lambda_{r}(g): f(h) \mapsto f(hg), \quad   \lambda_{l}(g): f(h)
  \mapsto f(g^{-1}h), \qquad \textrm{where } f(h)\in\FSpace{L}{2}(\Space{H}{n}).
\end{equation}

As any group \(\Space{H}{n}\) acts on itself by the conjugation automorphisms
\(\object[]{A}(g) h= g^{-1}hg\), which fix the unit \(e\in
\Space{H}{n}\). The differential \(\object[]{Ad}: \algebra{h}^n\rightarrow
\algebra{h}^n\) of \(\object[]{A}\) at \(e\) is a linear map which can
be differentiated again to the representation \(\object[]{ad}\) of the Lie algebra
\(\algebra{h}^n\) by the
commutator: \(\object{ad}(A): B \mapsto [B,A]\). The adjoint space
\(\algebra{h}^*_n\) of the Lie algebra 
\(\algebra{h}^n\) can be realised by the left invariant first order
differential forms on \(\Space{H}{n}\). By the duality
between \(\algebra{h}^n\) and \(\algebra{h}^*_n\) the map \(\object[]{Ad}\)
generates the \emph{co-adjoint
representation}~\cite[\S~15.1]{Kirillov76} \(\object[^*]{Ad}: \algebra{h}^*_n
\rightarrow \algebra{h}^*_n\):
\begin{equation}
  \label{eq:co-adjoint-rep}
  \object[^*]{Ad}(s,x,y): (\myhbar ,q,p) \mapsto (\myhbar , q+\myhbar y,
  p-\myhbar x), \quad
  \textrm{where } (s,x,y)\in \Space{H}{n} 
\end{equation}
and \((\myhbar ,q,p)\in\algebra{h}^*_n\) in bi-orthonormal coordinates to
the exponential ones on \(\algebra{h}^n\).  These coordinates \(h\),
\(q\), \(p\) should have units of an \emph{action} \(ML^2/T\), \emph{coordinates}
\(L\), and \emph{momenta} \(LM/T\) correspondingly. Again nothing
in~\eqref{eq:co-adjoint-rep} violates the Convention~\ref{conv:units-on-Hn}.

There are two types of orbits for
\(\object[^*]{Ad}\)~\eqref{eq:co-adjoint-rep}: isomorphic to Euclidean
spaces \(\Space{R}{2n}\) and single points:
\begin{eqnarray}
  \label{eq:co-adjoint-orbits-inf}
  \orbit{\myhbar} & = & \{(\myhbar, q,p): \textrm{ for a fixed
  }\myhbar\neq 0 \textrm{ and  all } (q,p) \in  \Space{R}{2n}\}, \\
  \label{eq:co-adjoint-orbits-one}
  \orbit{(q,p)} & = & \{(0,q,p): \textrm{ for a fixed } (q,p)\in \Space{R}{2n}\}.
\end{eqnarray} The \emph{orbit method} of
Kirillov~\cite[\S~15]{Kirillov76}, \cite{Kirillov99} starts from the
observation that the above orbits parametrise all irreducible unitary
representations of \(\Space{H}{n}\). All representations are
\emph{induced}~\cite[\S~13]{Kirillov76} by the character
\(\chi_\myhbar(s,0,0)=e^{2\pi\rmi \myhbar s}\) of the centre of
\(\Space{H}{n}\) generated by \((\myhbar,0,0)\in\algebra{h}^*_n\) and
shifts~\eqref{eq:co-adjoint-rep} from the ``left hand side'' (i.e. by \(g^{-1}\)) on
orbits. Using~\cite[\S~13.2, Prob.~5]{Kirillov76} we get a neat
formula, which (unlike some other in literature, e.g. \cite[Chap.~1,
(2.23)]{MTaylor86}) respects the Convention~\ref{conv:units-on-Hn} for all
physical units:
\begin{equation}
  \label{eq:stone-inf}
  \uir{\myhbar}(s,x,y): f_\myhbar (q,p) \mapsto 
  e^{-2\pi\rmi(\myhbar s+qx+py)}
  f_\myhbar \left(q-\frac{\myhbar}{2} y, p+\frac{\myhbar}{2} x\right).
\end{equation} Exactly the same formula is obtained if we apply the Fourier
transform \(\hat{\ }: \FSpace{L}{2}(\Space{H}{n})\rightarrow
\FSpace{L}{2}(\algebra{h}_n^*)\) given by:
\begin{equation}
  \label{eq:fourier-transform}
  \hat{\phi}(F)=\int_{\algebra{h}^n} \phi(\exp X) 
  e^{-2\pi\rmi  \scalar{X}{F}}\,dX \qquad \textrm{ where } X\in\algebra{h}^n,\ F\in\algebra{h}_n^*
\end{equation} to the left regular
action~\eqref{eq:left-right-regular}, see~\cite[\S~2.3]{Kirillov99}
for relations of the Fourier transform~\eqref{eq:fourier-transform}
and the orbit method. 

The derived representation \(d\uir{\myhbar}\) of  the Lie algebra
\(\algebra{h}^n\) defined on the vector
fields~\eqref{eq:h-lie-algebra} is:
\begin{equation} 
  d\uir{\myhbar}(S)=-2\pi\rmi  \myhbar I, \  
  d\uir{\myhbar}(X_j)=\frac{\myhbar}{2} \partial_{p_j}-2\pi\rmi
    q_j I, \  
  d\uir{\myhbar}(Y_j)=-\frac{\myhbar}{2} \partial_{q_j}-2\pi\rmi p_j I,
  \label{eq:der-repr-h-bar}
\end{equation} which clearly represents the commutation
rules~\eqref{eq:heisenberg-comm}.  The representation
\(\uir{\myhbar}\)~\eqref{eq:stone-inf} is reducible on whole
\(\FSpace{L}{2}(\orbit{\myhbar})\) as can be seen from the existence
of the set of ``right-invariant'', i.e. commuting
with~\eqref{eq:der-repr-h-bar}, differential operators:
\begin{equation} 
  \fl 
  d\uir{\myhbar}^r(S)=2\pi\rmi  \myhbar I, \quad 
  d\uir{\myhbar}^r(X_j)=-\frac{\myhbar}{2} \partial_{p_j}-
    2\pi\rmi  q_j I ,\quad 
  d\uir{\myhbar}^r(Y_j)=\frac{\myhbar}{2} \partial_{q_j}- 2\pi\rmi  p_j I,
  \label{eq:der-repr-h-bar-right}
\end{equation} 
which also represent the commutation
rules~\eqref{eq:heisenberg-comm}. 

To obtain an irreducible representation defined
by~\eqref{eq:stone-inf} we need to restrict it to a subspace of
\(\FSpace{L}{2}(\orbit{\myhbar})\) where operators
\eqref{eq:der-repr-h-bar-right} acts as scalars, e.g. use a
\emph{polarisation} from the \emph{geometric
quantisation}~\cite{Woodhouse80}.  For \(\myhbar>0\) consider  the
vector field \( -X_j+\rmi c_\rmi Y_j\) from the complexification of
\(\algebra{h}^n\), where the constant \(c_\rmi\) has the dimension
\(T/M\) in order to satisfy the Convention~\ref{conv:units-on-Hn}, the numerical
value of \(c_\rmi\) in given units can be assumed \(1\). We
introduce operators \(D^j_\myhbar\), \(1\leq j\leq n\) representing
vectors \(-X_j+\rmi c_\rmi Y_j\):
\begin{equation}
  \label{eq:Cauchy-Riemann}
  D^j_\myhbar =d\uir{\myhbar}^r(-X_j+\rmi c_\rmi Y_j)
  =\frac{\myhbar}{2}  (\partial_{p_j}+ c_\rmi \rmi \partial_{q_j})+
  2\pi(c_\rmi p_j+ \rmi   q_j) I
  =\myhbar \partial_{\bar{z}_j}+2\pi{z_j} I
\end{equation} where \( z_j = c_\rmi p_j+\rmi q_j\). For
\(\myhbar<0\) we define \(D^j_\myhbar =d\uir{\myhbar}^r(- c_\rmi
Y_j+\rmi X_j)\). 
Operators~\eqref{eq:Cauchy-Riemann} are used to give the
following classical result in terms of orbits:
\begin{thm}[Stone--von Neumann,
  cf. \textup{\cite[Chap.~1, \S~5]{Folland89}, \cite[\S~18.4]{Kirillov76}}]
  \label{th:Stone-von-Neumann} 
  All unitary irreducible representations of \(\Space{H}{n}\) are
  parametrised up to equivalence by two classes of
  orbits~\eqref{eq:co-adjoint-orbits-inf}
  and~\eqref{eq:co-adjoint-orbits-one} of co-adjoint
  representation~\eqref{eq:co-adjoint-rep} in \(\algebra{h}^*_n\): 
  \begin{enumerate}
  \item The infinite dimensional representations by transformation
    \(\uir{\myhbar}\)~\eqref{eq:stone-inf} for \(\myhbar \neq 0\) in
    Fock~\textup{\cite{Folland89,Howe80b}} space
    \(\FSpace{F}{2}(\orbit{\myhbar})\subset\FSpace{L}{2}(\orbit{\myhbar})\)
    of null solutions to the operators \(D^j_\myhbar\)
    \eqref{eq:Cauchy-Riemann}:
    \begin{equation}
      \label{eq:Fock-type-space}
      \FSpace{F}{2}(\orbit{\myhbar})=\{f_{\myhbar}(q,p) \in
      \FSpace{L}{2}(\orbit{\myhbar}) \such D^j_\myhbar f_{\myhbar}=0,\
       1 \leq j \leq n\}.
    \end{equation}
  \item The one-dimensional representations as multiplication by
    a constant on \(\Space{C}{}=\FSpace{L}{2}(\orbit{(q,p)})\) which
    drop out from~\eqref{eq:stone-inf} for \(\myhbar =0\):
    \begin{equation}
      \label{eq:stone-one}
      \uir{(q,p)}(s,x,y): c \mapsto e^{-{2\pi \rmi }(qx+py)}c,
    \end{equation}
    with the corresponding derived representation
    \begin{equation} 
      \fl 
      d\uir{(q,p)}(S)=0, \qquad 
      d\uir{(q,p)}(X_j)=-2\pi\rmi q_j ,\qquad
      d\uir{(q,p)}(Y_j)=-2\pi\rmi p_j.
      \label{eq:der-repr-qp}
    \end{equation} 
  \end{enumerate}
\end{thm}

\subsection{Structure and Topology of the Unitary Dual of
  $\Space{H}{n}$}
\label{sec:struct-topol-unit} 
\begin{figure}[tbp]
  \begin{center}
    \includegraphics{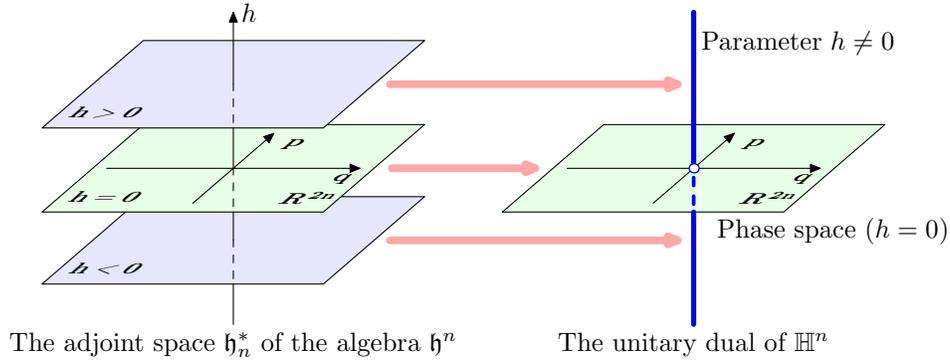}
  \end{center}
  \caption[The unitary dual object to the Heisenberg
  group]{The structure of unitary dual object to \(\Space{H}{n}\)
    appearing from the method of orbits. The space \(\algebra{h}_n^*\)
    is sliced into ``horizontal'' hyperplanes. Planes with
    \(\myhbar\neq0\) form single orbits and correspond to different
    classes of unitary irreducible representation. The plane \(\myhbar
    =0\) is a family of one-point orbits \((0,q,p)\), which produce
    one-dimensional representations. The topology on the dual object
    is the factor topology inherited from the
    \(\algebra{h}_n^*\)~\cite[\S~2.2]{Kirillov99}.}
  \label{fi:unitary-dual}
\end{figure}

The structure of the unitary dual object
to \(\Space{H}{n}\)---the collection of all different classes of
unitary irreducible representations---as it appears from the method of
orbits is illustrated by the Figure~\ref{fi:unitary-dual},
cf.~\cite[Chap.~7, Fig.~6 and 7]{Kirillov94a}. The adjoint space
\(\algebra{h}_n^*\) is sliced into ``horizontal'' hyperplanes. A plane
with a parameter \(\myhbar\neq0\) forms a single
orbit~\eqref{eq:co-adjoint-orbits-inf} and corresponds to a particular
class of unitary irreducible representation~\eqref{eq:stone-inf}. The
plane with parameter \(\myhbar =0\) is a family of one-point orbits
\((0,q,p)\)~\eqref{eq:co-adjoint-orbits-one}, which produce
one-dimensional representations~\eqref{eq:stone-one}. The topology on
the dual object is the factor topology inherited from the adjoint
space \(\algebra{h}_n^*\) under the above identification,
see~\cite[\S~2.2]{Kirillov99}.
\begin{example}
  \label{ex:density-of-representations} A set of representations
  \(\uir{\myhbar}\)~\eqref{eq:stone-inf} with \(\myhbar\rightarrow 0\)
  is dense in the whole family of one-dimensional
  representations~\eqref{eq:stone-one}, as can be seen either from
  the Figure~\ref{fi:unitary-dual} or the analytic
  expressions~\eqref{eq:stone-inf} and~\eqref{eq:stone-one} for those
  representations.
\end{example}
 Non-commutative representations \(\uir{\myhbar}\),
\(\myhbar\neq0\)~\eqref{eq:stone-inf} are known to be connected with
quantum mechanics~\cite{Folland89} from its origin.  This explains,
for example, the name of the Heisenberg group. In the contrast
commutative representations~\eqref{eq:stone-one} are mostly neglected
and only mentioned for sake of completeness in some mathematical
formulations of the Stone--von Neumann theorem. The development of
\emph{\(p\)-mechanics} started~\cite{Kisil96a} from the observation
that the union of all representations \(\uir{(q,p)}\),
\((q,p)\in\Space{R}{2n}\) naturally acts as the classical \emph{phase
space}. The sensibleness of the single union
\begin{equation}
  \label{eq:orbit-0}
  \orbit{0}=\bigcup_{(q,p)\in\Space{R}{2n}} \orbit{(q,p)}
\end{equation}
rather than unrelated set of disconnected orbits 
manifests itself in several ways:
\begin{enumerate}
\item The topological position of \(\orbit{0}\) as the limiting case
  (cf. Example~\ref{ex:density-of-representations}) of quantum
  mechanics for \(\myhbar\rightarrow 0\) realises the
  \emph{correspondence principle} between quantum and classical
  mechanics.
\item Symplectic automorphisms of the Heisenberg group (see
  Subsection~\ref{sec:quant-from-sypl}) produce the metaplectic
  representation in quantum mechanics and \emph{transitively} act by
  linear symplectomorphisms on the whole set \(\orbit{0}\setminus \{0\}\).
\item We got the Poisson brackets~\eqref{eq:Poisson} on \(\orbit{0}\)
  from the same source~\eqref{eq:u-star} that leads to the correct
  Heisenberg equation in quantum mechanics.
\end{enumerate}

The identification of \(\orbit{0}\) with the classical phase space
justifies that \(q\) and \(p\) are measured by the units of length and
momentum respectively, which supports our choice of units for \(x\) and
\(y\) in Convention~\ref{it:a-priori}.

\begin{rem}
  \label{re:equivalence} Since unitary representations are classified
  up to a unitary equivalence one may think that their explicit
  realisations in particular Hilbert spaces are ``the same''. However
  a suitable form of a representation can give many technical
  advantages. The classical illustration is the paper~\cite{Howe80b},
  where the comparison of the (unitary equivalent!)  Schr\"odinger and
  Fock representations of \(\Space{H}{n}\) is the principal tool of
  investigation.
\end{rem}

Our form~\eqref{eq:stone-inf} of representations of \(\Space{H}{n}\)
given in Theorem~\ref{th:Stone-von-Neumann} has at least two following
advantages, which are rarely combined together:
\begin{enumerate}
\item There is the explicit physical meaning of all entries
  in~\eqref{eq:stone-inf} as will be seen bellow. In the contrast the
  formula (2.23) in~\cite[Chap.~1]{MTaylor86} contains terms
  \(\sqrt{\myhbar}\) (in our notations), which could be hardly justified
  from a physical point of view.
\item The one-dimension representations~\eqref{eq:stone-one}
  explicitly correspond to the case \(\myhbar=0\)
  in~\eqref{eq:stone-inf}. The Schr\"odinger representation (the most
  used in quantum mechanics!) is handicapped in this sense: a
  transition \(\myhbar\rightarrow 0\) from \(\uir{\myhbar}\) in
  the Sch\"odinger form to \(\uir{(q,p)}\) requires a long
  discussion~\cite[Ex.~7.11]{Kirillov94a}.
\end{enumerate}

We finish the discussion of the unitary dual of \(\Space{H}{n}\) by a
remark about negative values of \(\myhbar\). Since its position in the
Heisenberg equation~\eqref{eq:moyal-equation} a negative value of
\(\hbar\)  revert the flow of time. Thus representations
\(\uir{\myhbar}\) with \(\myhbar<0\) seems to be suitable for
a description of anti-particles. There is the explicit
(cf. Figure~\ref{fi:unitary-dual}) mirror symmetry between matter and
anti-matter through classical mechanics. In this paper however we
will consider only the case of \(\myhbar>0\).

\subsection{Fock Spaces $\FSpace{F}{2}(\orbit{\myhbar})$ and
  Coherent States}
\label{sec:fock-type-spaces}
Our Fock type spaces~\eqref{eq:Fock-type-space} are not very
different~\cite[Ex.~4.3]{Kisil98a} from the standard Segal--Bargmann
spaces. 
\begin{defn} \textup{\cite{Folland89,Howe80b}}
  The Segal--Bargmann space (with a parameter \(\myhbar>0\)) consists
  of functions on \(\Space{C}{n}\) which are holomorphic in \(z\),
  i.e. \(\partial_{\bar{z}_j} f(z)=0\),  and
  square integrable with respect to the measure \(e^{-2\modulus{z}^2/\myhbar
  }dz\) on \(\Space{C}{n}\):
  \begin{displaymath}
    \int_{\Space{C}{n}} \modulus{f(z)}^2e^{-2\modulus{z}^2/\myhbar }dz < \infty.
  \end{displaymath}
\end{defn}
Noticing the \(\partial_{\bar{z}_j}\) component in the operator
\(D^j_\myhbar\)~\eqref{eq:Cauchy-Riemann} we obviously obtain
\begin{prop}
 A function
  \(f_\myhbar(q,p)\) is in
  \(\FSpace{F}{2}(\orbit{\myhbar})\)~\eqref{eq:Fock-type-space} for
  \(\myhbar>0\) if and only if the function
  \(f_\myhbar(z)e^{\modulus{z}^2/\myhbar}\), \(z=p+\rmi q\) is in
  the classical Segal--Bargmann space.
\end{prop}

The space \(\FSpace{F}{2}(\orbit{\myhbar})\) can be also described in
the language of \emph{coherent states}, (also known as
\emph{wavelets}, \emph{matrix elements} of representation,
\emph{Berezin transform}, etc., see~\cite{AliAntGaz00,Kisil98a}).
Since the representation \(\uir{\myhbar}\) is irreducible any vector
\(v_0\) in \(\FSpace{F}{2}(\orbit{\myhbar})\) is \emph{cyclic},
i.e. vectors \(\uir{\myhbar}(g)v_0\) for all \(g\in G\) span the whole
space \(\FSpace{F}{2}(\orbit{\myhbar})\). Even all vectors
are equally good in principle, some of them are more equal for
particular purposes (cf. Remark~\ref{re:equivalence}). For the
harmonic oscillator the preferred vector is the dimensionless
\emph{vacuum state}:
\begin{equation}
  \label{eq:vacuum}
  v_0(q,p)= \exp\left(-\frac{2\pi}{\myhbar}\left( c_\rmi^{-1} q^2
    + c_\rmi p^2\right)\right), 
\end{equation} which corresponds to the minimal level of energy. Here
\(c_\rmi\) as was defined before~\eqref{eq:Cauchy-Riemann} has the
dimensionality \(T/M\). One can check directly the validity of the
both equation~\eqref{eq:Fock-type-space} and Convention~\ref{conv:units-on-Hn}
for~\eqref{eq:vacuum}, particularly the exponent is taken from a
dimensionless pure number. Note also that \(v_0(q,p)\) is destroyed by
the \emph{annihilation operators} (sf.~\eqref{eq:der-repr-h-bar}
and~\eqref{eq:Cauchy-Riemann}):
\begin{equation}
  \label{eq:annihilation-operator}
  A_\myhbar^j=d\uir{\myhbar}( X_j+\rmi c_\rmi Y_j)=\frac{\myhbar}{2}  (\partial_{p_j}- \rmi c_\rmi
  \partial_{q_j})+2\pi(c_\rmi  p_j- \rmi q_j) I.
\end{equation}

We introduce a dimensionless inner product on \(\FSpace{F}{}(\orbit{\myhbar})\) by
the formula:
\begin{equation}
  \label{eq:inner-product}
  \scalar{f_1}{f_2}=\left(\frac{4}{\myhbar}\right)^{n}
  \int_{\Space{R}{2n}} f_1(q,p)\,\bar{f}_2(q,p)\,dq\,dp
\end{equation} With respect to this product the vacuum
vector~\eqref{eq:vacuum} is normalised: \(\norm{v_0}=1\). 
For any observable \(A\) the formula
\begin{displaymath}
  \scalar{A v_0}{v_0}=\left(\frac{4}{\myhbar}\right)^{n} 
  \int_{\Space{R}{2n}} Av_0(q,p)\,\bar{v}_0(q,p)\,dq\,dp
\end{displaymath} gives an expectation in the units of \(A\) since
both the vacuum vector \(v_0(q,p)\) and the inner
product~\eqref{eq:inner-product} are dimensionless. The term
\(\myhbar^{-n}\) in~\eqref{eq:inner-product} does not only normalise the
vacuum and fix the dimensionality of the inner product; it is also
related to the \emph{Plancherel measure}~\cite[(1.61)]{Folland89},
\cite[Chap.~1, Th.~2.6]{MTaylor86} on the unitary dual of
\(\Space{H}{n}\).

Elements \((s,0,0)\) of the centre of \(\Space{H}{n}\) trivially act
in the representation \(\uir{\myhbar}\)~\eqref{eq:stone-inf} as
multiplication by scalars,
e.g. any function is a common eigenvector of all operators
\(\uir{\myhbar}(s,0,0)\). Thus the essential
part~\cite[Defn.~2.5]{Kisil98a} of the operator
\(\uir{\myhbar}(s,x,y)\) is determined solely by
\((x,y)\in\Space{R}{2n}\). The \emph{coherent states}
\(v_{(x,y)}(q,p)\) are ``left shifts'' of the vacuum vector
\(v_0(q,p)\) by operators~\eqref{eq:stone-inf}:
\begin{eqnarray}
  \label{eq:coherent-states}
  v_{(x,y)}(q,p) &=&\uir{\myhbar}(0,x,y) v_0(q,p) \\
  &=& \exp\left(-2\pi\rmi(qx+py)-\frac{2\pi}{\myhbar} \left(
      c_\rmi^{-1}\left(q-\frac{\myhbar}{2} y\right)^
      2+c_\rmi \left(p+\frac{\myhbar}{2} x\right)^2\right)\right). \nonumber 
\end{eqnarray}

Now any function from the space \(\FSpace{F}{2}(\orbit{\myhbar})\)
can be represented~\cite[Ex.~4.3]{Kisil98a} as a linear superposition of
coherent states: 
\begin{eqnarray}
  \label{eq:inv-wavelet-transform}
  f(q,p)=[\oper{M}_\myhbar \breve{f}](q,p)&=& \myhbar^{n} 
  \int_{\Space{R}{2n}} \breve{f}(x,y)
  v_{(x,y)}(q,p)\,dx\,dy\\
  &=& \myhbar^{n} \int_{\Space{R}{2n}} \breve{f}(x,y)
  \uir{\myhbar}(x,y)\,dx\,dy\, v_{(0,0)}(q,p) \nonumber
\end{eqnarray}
where \(\breve{f}(x,y)\) is the \emph{wavelet} (or \emph{coherent states})
  \emph{transform}~\cite{AliAntGaz00,Kisil98a} of \(f(q,p)\): 
\begin{eqnarray}
  \label{eq:wavelet-transform}
  \breve{f}(x,y) =[\oper{W}_\myhbar f](x,y)
  &=&\scalar{f}{v_{(x,y)}}_{\FSpace{F}{2}(\orbit{\myhbar})}\\
  &=& \left(\frac{4}{\myhbar}\right)^{n}  \int_{\Space{R}{2n}}
  f(q,p)\bar{v}_{(x,y)}(q,p)\,dq\,dp. \nonumber 
\end{eqnarray} The formula~\eqref{eq:inv-wavelet-transform} can be
regarded~\cite{Kisil98a} as the \emph{inverse wavelet transform}
\(\oper{M}\) of \(\breve{f}(x,y)\).  Note that all above integrals are
dimensionless, thus both the wavelet transform and its inverse are
measured in the same units.

The straightforward use of the basic formula:
\begin{equation}
  \label{eq:axx-int}
  \int\limits_{-\infty}^\infty
  \exp(-ax^2+bx+c)\,dx 
  =\sqrt{\frac{\pi}{a}}\exp\left(\frac{b^2}{4a}+c\right), \qquad \textrm{ where } a>0.
\end{equation} for the wavelet
transform~\eqref{eq:inv-wavelet-transform} leads to:
\begin{equation}
  \label{eq:v0-transform}
  \breve{v}_0(s,x,y)= \exp 2\pi\left(
    \rmi \myhbar s -\frac{\myhbar}{4}\left(c_\rmi x^2+c_\rmi^{-1} y^2\right)\right).
\end{equation} Since~\cite[Prop.~2.6]{Kisil98a} the wavelet transform
\(\oper{W}_\myhbar\)~\eqref{eq:inv-wavelet-transform} intertwines
\(\uir{\myhbar}\)~\eqref{eq:stone-inf} with the left regular
representation \(\lambda_l\)~\eqref{eq:left-right-regular}:
\begin{displaymath}
  \oper{W}_\myhbar \circ \uir{\myhbar}(g)= \lambda_l(g)\circ \oper{W}_\myhbar,
  \qquad \textrm{ for all } g\in\Space{H}{n} 
\end{displaymath} the image of an arbitrary coherent state is:
\begin{eqnarray}
  \lefteqn{\breve{v}_{(s',x',y')} (s,x,y)=}    \label{eq:W-cogerent-states} \\
&&\exp 2\pi\left(
    \rmi \myhbar \left(s-s'-\frac{1}{2}(x'y-xy')\right)
    -\frac{\myhbar}{4}\left(c_\rmi (x-x')^2+c_\rmi^{-1}
      (y-y')^2\right)\right).
  \nonumber 
\end{eqnarray}
Needless to say that these functions  are obeying
Convention~\ref{conv:units-on-Hn}. 

We should mention however a problem related to coherent
states~\eqref{eq:coherent-states}: all their ``classical limits'' for
\(\myhbar\rightarrow 0\) are functions with supports in neighbourhoods
of \((0,0)\). In the contrast we may wish them be supported around
different classical states \((q,p)\). This difficulty can be resolved
through a replacement of the group action of \(\Space{H}{n}\)
in~\eqref{eq:coherent-states} by the
``shifts''~\eqref{eq:shifts-on-orbits} generated by the
\(p\)-mechanical brackets~\eqref{eq:u-brackets}.

\section{$p$-Mechanics: Statics}
\label{sec:p-mechanics-statics}
We define \(p\)-mechanical observables to be convolutions on the
Heisenberg group. The next Subsection describes their multiplication
and commutator as well as quantum and classical representations. The
Berezin quantisation in form of wavelet transform is considered in
Subsection~\ref{sec:berezin-quantisation}. This is developed in
Subsection~\ref{sec:weyl-quantisation} into a construction of
\(p\)-observables out of either quantum or classical ones.

\subsection{Observables in $p$-Mechanics, Convolutions and Commutators}
\label{sec:conv-algebra-hg}

In line with the standard quantum theory we give the following definition:
\begin{defn}
  \label{de:p-observables}
  \emph{Observables} in \(p\)-mechanics
  (\(p\)-observables) are
  presented by operators on \(\FSpace{L}{2}(\Space{H}{n})\).
\end{defn}
Actually we will need here\footnote{More
  general operators are in use for a string-like version of
  \(p\)-mechanics, see Subsection~\ref{sec:string-theory}.} only operators
generated by convolutions on \(\FSpace{L}{2}(\Space{H}{n})\).  Let
\(dg\) be a left invariant measure~\cite[\S~7.1]{Kirillov76} on
\(\Space{H}{n}\), which coincides with the standard Lebesgue measure
on \(\Space{R}{2n+1}\) in the exponential coordinates
\((s,x,y)\). Then a function \(k_1\) from the linear space
\(\FSpace{L}{1}(\Space{H}{n},dg)\) acts on
\(k_2\in\FSpace{L}{2}(\Space{H}{n},dg)\) by the convolution as
follows:
\begin{eqnarray}
  (k_1 * k_2) (g)   &=& c_\myhbar^{n+1}
  \int_{\Space{H}{n}} k_1(g_1)\,
  k_2(g_1^{-1}g)\,dg_1 \label{eq:de-convolution}\\
  &=&   c_\myhbar^{n+1}
  \int_{\Space{H}{n}} k_1(gg_1^{-1})\,
  k_2(g_1)\,dg_1,
  \nonumber
\end{eqnarray} where the constant \(c_\myhbar\) is measured in units
of the action and can be assumed equal to \(1\). Then
\(c_\myhbar^{n+1}\) has units inverse to \(dg\). Thus the convolution
\(k_1 * k_2\) is measured in units those are product of units for
\(k_1\) and \(k_2\).  The composition of two convolution operators
\(K_1\) and \(K_2\) with kernels \(k_1\) and \(k_2\) has the kernel
defined by the same formula~\eqref{eq:de-convolution}. Clearly two
products \(K_1K_2\) and \(K_2K_1\) could have a different value due to
non-commutativity of \(\Space{H}{n} \) but always are measured in the
same units. Thus we can find out how distinct they are from the
difference \(K_1K_2-K_2K_1\), which does not violate the
Convention~\ref{conv:units-on-Hn}. This also produces the \emph{inner
derivations} \(D_k\) of \(\FSpace{L}{1}(\Space{H}{n})\) by the
\emph{commutator}:
\begin{eqnarray}
  D_k: f \mapsto [k,f]&=&k*f-f*k \nonumber \\
  &=& c_\myhbar^{n+1}
  \int_{\Space{H}{n}} k(g_1)\left( f(g_1^{-1}g)-f(gg_1^{-1})\right)\,dg_1.
  \label{eq:commutator}
\end{eqnarray} Because we only consider observables those are
convolutions on \(\Space{H}{n}\) we can extend a unitary
representation \(\uir{\myhbar} \) of \(\Space{H}{n}\) to a
\(*\)-representation \(\FSpace{L}{1}(\Space{H}{n} ,dg)\) by the
formula:
\begin{eqnarray}
  \fl
  [\uir{\myhbar} (k)f](q,p)
 & =&  c_\myhbar^{n+1} \int_{\Space{H}{n}} k(g)\uir{\myhbar}
  (g)f(q,p)\,dg  \nonumber \\
  &=& c_\myhbar^{n} \int_{\Space{R}{2n}}
  \left(c_\myhbar\int_{\Space{R}{}} k(s,x,y) 
    e^{-2\pi\rmi \myhbar s}\,ds \right)  \label{eq:rho-extended-to-L1} \\
  && \qquad \qquad \times e^{- 2\pi\rmi (qx+py)}      
  f \left(q-\frac{\myhbar}{2} y, p+\frac{\myhbar}{2} x\right) \,dx\,dy.
  \nonumber
\end{eqnarray} The last formula in the Schr\"odinger representation
defines for \(\myhbar \neq 0\) a \emph{pseudodifferential
  operator}~\cite{Folland89,Howe80b,Shubin87}  on
\(\FSpace{L}{2}(\Space{R}{n})\)~\eqref{eq:Fock-type-space}, which
are known to be \emph{quantum observables} in the Weyl quantisation. For
representations \(\uir{(q,p)}\)~\eqref{eq:stone-one} an expression
analogous to~\eqref{eq:rho-extended-to-L1} defines an operator of
multiplication on \(\orbit{0}\)~\eqref{eq:orbit-0} by the Fourier
transform of \(k(s,x,y)\):
\begin{equation}
  \label{eq:classical-observables}
  \uir{(q,p)}(k)=\hat{k}\left(0,{q},{p}\right)
  =  c_\myhbar^{n+1} \int_{\Space{H}{n}}
  k(s,x,y)e^{-{2\pi\rmi}(qx+py)} \,ds\,dx\,dy,
\end{equation} where the direct \(\hat{\ }\) and inverse \(\check{\
}\) Fourier transforms are defined by the formulae:
\begin{displaymath}
  \hat{f}(v)=\int_{\Space{R}{m}} f(u)
  e^{-2\pi\rmi uv}\,du \qquad \textrm{ and } \qquad
  f(u)=(\hat{f})\check{\ }(u)=\int_{\Space{R}{m}} \hat{f}(v)
  e^{2\pi\rmi vu}\,dv.
\end{displaymath} For reasons discussed in
subsections~\ref{sec:struct-topol-unit}
and~\ref{sec:p-mechanical-bracket} we regard the
functions~\eqref{eq:classical-observables} on \(\orbit{0}\) as
\emph{classical observables}.  Again the both representations
\(\uir{\myhbar} (k)\) and \(\uir{(q,p)}k\) are measured in the same
units as the function \(k\) does.

From~\eqref{eq:rho-extended-to-L1} follows that \(\uir{\myhbar}
(k)\) for a fixed \(\myhbar \neq 0\) depends only from
\(\hat{k}_s(\myhbar,x,y)\), which is the partial Fourier transform
\(s\rightarrow \myhbar\) of \(k(s,x,y)\). Then the representation of
the composition of two convolutions depends only from
\begin{eqnarray}
  \fl
  (k'*k)\hat{_s}
  &=& 
  c_\myhbar \int_{\Space{R}{}}e^{ -2\pi\rmi \myhbar s}\,
  c_\myhbar^{n+1} \int_{\Space{H}{n}}k'(s',x',y') 
 \label{eq:almost-star-product}\\
 &&\quad \qquad \times k(s-s'+\frac{1}{2}
 (xy'-yx'),x-x',y-y')\,ds'dx'dy'ds \nonumber \\
  &=&    c_\myhbar^{n} \int_{\Space{R}{2n}}e^{{\pi\rmi \myhbar} (xy'-yx')}
   \cdot c_\myhbar \int_{\Space{R}{}} e^{ -2\pi\rmi    \myhbar s'} 
  k'(s',x',y') \,ds' \nonumber \\
  && \qquad \qquad
  \times c_\myhbar \int_{\Space{R}{}}e^{ -2\pi\rmi \myhbar (s-s'+\frac{1}{2}
 (xy'-yx'))}\nonumber \\
 &&\qquad \qquad \quad \times k(s-s'+\frac{1}{2}
 (xy'-yx'),x-x',y-y')\, ds\,dx'dy' \nonumber \\
  &=&  c_\myhbar^{n}
  \int_{\Space{R}{2n}} e^{ {\pi\rmi \myhbar}{}
   (xy'-yx')} \hat{k}'_s(\myhbar ,x',y')
 \hat{k}_s(\myhbar ,x-x',y-y')\,dx'dy'. \nonumber 
\end{eqnarray} Note that if we apply the Fourier transform
\((x,y)\rightarrow (q,p)\) to the last
expression~\eqref{eq:almost-star-product} then we get the \emph{star
product} of \(\hat{k}'\) and \(\hat{k}\) known in \emph{deformation}
quantisation, cf. \cite[(9)--(13)]{Zachos02a}.  Consequently the
representation \(\uir{\myhbar}([k',k])\) of the
commutator~\eqref{eq:commutator} depends only from:
\begin{eqnarray}
  \fl
\label{eq:repres-commutator}
  \lefteqn{[k',k]\hat{_s}
  =
   c_\myhbar^{n}\int_{\Space{R}{2n}}   
  \left(e^{ {\rmi\pi}{} \myhbar  (xy'-yx')}-e^{- {\rmi\pi}{} \myhbar  (xy'-yx')}\right)}
  \qquad \\
  && \qquad \qquad \qquad \times 
  \hat{k}'_s(-\myhbar ,x',y')
  \hat{k}_s(-\myhbar ,x-x',y-y') \,dx'dy'\nonumber \\
 \qquad&=&    2 \rmi c_\myhbar^n  \int_{\Space{R}{2n}}\! \sin\left({\pi\myhbar}
   (xy'-yx')\right)
 \hat{k}'_s(\myhbar ,x',y')
 \hat{k}_s(\myhbar ,x-x',y-y')\,dx'dy'.\nonumber 
\end{eqnarray} The integral~\eqref{eq:repres-commutator} turns to be
equivalent to the \emph{Moyal brackets}~\cite{Zachos02a} for the
(full) Fourier transforms of \(k'\) and \(k\). It is commonly accepted
that the method of orbit is a mathematical side of the \emph{geometric}
quantisation~\cite{Woodhouse80}. Our derivation of the Moyal brackets
in terms of orbits shows that deformation and geometric quantisations
are closely connected and both are not very far from the original
quantisation of Heisenberg and Schr\"odinger. Yet one more their close
relative can be identified as the Berezin
quantisation~\cite{Berezin75b}, see the next Subsection.

\begin{rem}
  \label{re:triv-commutator}
  The expression~\eqref{eq:repres-commutator} vanishes for
  \(\myhbar=0\) as can be expected from the commutativity of
  representations~\eqref{eq:stone-one}. Thus it does not produce
  anything interesting on \(\orbit{0}\), that supports the common
  negligence to this set.
\end{rem}

Summing up, \(p\)-mechanical observables, i.e. convolutions on
\(\FSpace{L}{2}(\Space{H}{n})\), are transformed
\begin{enumerate}
\item  by representations   \(\uir{\myhbar}\)~\eqref{eq:stone-inf}
  into quantum observables~\eqref{eq:rho-extended-to-L1} with the
  Moyal bracket~\eqref{eq:repres-commutator} between them;
\item by representations \(\uir{(q,p)}\)~\eqref{eq:stone-one} into
    classical observables~\eqref{eq:classical-observables}.
\end{enumerate}
We did not get a meaningful brackets on classical observables yet, this
will be done in Section~\ref{sec:p-mechanical-bracket}.

\subsection{Berezin Quantisation and Wavelet Transform}
\label{sec:berezin-quantisation}

There is the following construction, known as the \emph{Berezin
  quantisation}~\cite{Berezin72,Berezin75b},
allowing us to assign a function to an operator (observable) and an
operator to a function. The scheme is based on the construction of the
\emph{coherent states} and can be derived from different
sources~\cite{Klauder94b,Perelomov86}. We prefer the group-theoretic
origin of Perelomov coherent states~\cite{Perelomov86}, which is
realised in~\eqref{eq:coherent-states}. Following~\cite{Berezin72} we
introduce the \emph{covariant} symbol \({a}(g)\) of an operator \(A\) on
\(\FSpace{F}{2}(\orbit{\myhbar})\) by the simple expression:
\begin{equation}
  \label{eq:covariant}
  {a}(g)=\scalar{A v_g}{v_g},
\end{equation}
i.e. we get a map from the linear space of operators on
\(\FSpace{F}{2}(\orbit{\myhbar})\) to a linear space of function on \(\Space{H}{n}\).
A map in the opposite direction assigns to a function \(\breve{a}(g)\) on
\(\Space{H}{n}\) the linear operator \(A\) on
\(\FSpace{F}{2}(\orbit{\myhbar})\) by the formula
\begin{equation}
  \label{eq:contravariant}
  A=c_\myhbar^{n+1}\int_{\Space{H}{n}} \circstack{a}(g)P_g\,dg, \qquad \textrm{ where \(P_g\) is
    the projection } P_g
  v=\scalar{v}{v_g}v_g. 
\end{equation}
The function \(\circstack{a}(g)\) is called the  \emph{contravariant} symbol of
the operator \(A\)~\eqref{eq:contravariant}. 

The co- and contravariant symbols of operators are defined through the
coherent states, in fact both types of symbols are
realisations~\cite[\S~3.1]{Kisil98a} of the
direct~\eqref{eq:wavelet-transform} and
inverse~\eqref{eq:inv-wavelet-transform} wavelet transforms.  Let us
define a representation \(\uir{b\myhbar}\) of the group \(\Space{H}{n}
\times \Space{H}{n} \) in the space
\(\oper{B}(\FSpace{F}{2}(\orbit{\myhbar}))\) of operators on
\(\FSpace{F}{2}(\orbit{\myhbar})\) by the formula:
\begin{equation}
  \label{eq:bi-representation}
  \uir{b\myhbar}(g_1,g_2): A \mapsto
  \uir{\myhbar}(g_1^{-1})A\uir{\myhbar}(g_2), \qquad 
  \textrm{ where } g_1,g_2\in\Space{H}{n}.
\end{equation}
According to the scheme from~\cite{Kisil98a} for any state \(f_0\) on
\(\oper{B}(\FSpace{F}{2}(\orbit{\myhbar}))\) we get the wavelet
transform \(\oper{W}_{f_0}: \oper{B}(\FSpace{F}{2}(\orbit{\myhbar}))\rightarrow
\FSpace{C}{}(\Space{H}{n}\times\Space{H}{n}) \):
\begin{equation}
  \label{eq:wavelet-transform-operators}
  \oper{W}_{f_0}: A \mapsto \breve{a}(g_1,g_2)= \scalar{\uir{b\myhbar}(g_1,g_2)A}{f_0}.
\end{equation} 

The important particular case is given by \(f_0\)
defined through the vacuum vector \(v_0\)~\eqref{eq:vacuum} by the
formula \(\scalar[\oper{B}(\FSpace{F}{2}(\orbit{\myhbar}))]{A}{f_0} 
=\scalar[\FSpace{F}{2}(\orbit{\myhbar})]{Av_0}{v_0}\). Then the wavelet
transform~\eqref{eq:wavelet-transform-operators} produces the
\emph{covariant presymbol} \(\breve{a}(g_1,g_2)\) of operator
\(A\). Its restriction \(a(g)=\breve{a}(g,g)\) to the diagonal \(D\) of
\(\Space{H}{n}\times\Space{H}{n}\) is exactly~\cite{Kisil98a} the
Berezin covariant symbol~\eqref{eq:covariant} of~\(A\). Such a
restriction to the diagonal is done without a lost of information due
to holomorphic properties of
\(\breve{a}(g_1,g_2)\)~\cite{Berezin72}.  

Another important example of the state \(f_0\) is given by the trace:
\begin{equation}
  \label{eq:trace}
  \scalar{A}{f_0}=\object{tr}A
  =\myhbar^n \int_{\Space{R}{2n}}
  \scalar[\FSpace{F}{2}(\orbit{\myhbar})]{A
    v_{(x,y)}}{v_{(x,y)}}\,dx\,dy, 
\end{equation} where coherent states \(v_{(x,y)}\) are again defined
in~\eqref{eq:coherent-states}. Operators \(\uir{b\myhbar}(g,g)\) from
the diagonal \(D\) of \(\Space{H}{n}\times\Space{H}{n}\) trivially act
on the wavelet transform~\eqref{eq:wavelet-transform-operators}
generated by the trace~\eqref{eq:trace} since the trace is invariant
under \(\uir{b\myhbar}(g,g)\). According to the general
scheme~\cite{Kisil98a} we can consider \emph{reduced wavelet
transform} to the homogeneous space
\(\Space{H}{n}\times\Space{H}{n}/D\) instead of the entire group
\(\Space{H}{n}\times\Space{H}{n}\). The space
\(\Space{H}{n}\times\Space{H}{n}/D\) is isomorphic to \(\Space{H}{n}\)
with the embedding \(\Space{H}{n} \rightarrow \Space{H}{n} \times
\Space{H}{n}\) given by \(g \mapsto (g;0)\). Furthermore the centre
\(Z\) of \(\Space{H}{n}\) acts trivially in the representation
\(\uir{b\myhbar}\) as usual. Thus the only essential part of
\(\Space{H}{n}\times\Space{H}{n}/D\) in the wavelet transform is the
homogeneous space \(\Omega=\Space{H}{n}/Z\). A Borel section
\(\mathbf{s}: \Omega \rightarrow \Space{H}{n}\times\Space{H}{n} \) in
the principal bundle \(G \rightarrow \Omega\) can be defined as
\(\mathbf{s}: (x,y)\mapsto ((0,x,y);(0,0,0))\). We got the reduced
realisation \(\oper{W}_r\) of the wavelet
transform~\eqref{eq:wavelet-transform-operators} in the form:
\begin{eqnarray}
  \oper{W}_r: A \mapsto
  \breve{a}_r(x,y)&=&\scalar{\uir{b\myhbar}(\mathbf{s}(x,y)) A}{f_0}
  \nonumber\\
  &=&  \object{tr}(\uir{\myhbar}((0,x,y)^{-1}) A)
   \label{eq:inversion-trace} \\ 
  &=& \myhbar^{n} \int_{\Space{R}{2n}}
  \scalar[\FSpace{F}{2}(\orbit{\myhbar})]{
    \uir{\myhbar}((0,x,y)^{-1})A
    v_{(x',y')}}{v_{(x',y')}}\,dx'dy'
  \nonumber \\
  &=& \myhbar^{n} \int_{\Space{R}{2n}} \scalar[\FSpace{F}{2}(\orbit{\myhbar})]{A
    v_{(x',y')}}{v_{(x,y)\cdot(x',y')}}\,dx'dy'.
  \label{eq:inversion-formula}
\end{eqnarray} The formula~\eqref{eq:inversion-trace} is the principal
ingredient of the \emph{inversion formula} for the Heisenberg
group~\cite[Chap.~1, (1.60)]{Folland89}, \cite[Chap.~1,
Th.~2.7]{MTaylor86}, which reconstructs kernels of convolutions
\(k(g)\) out of operators \(\uir{\myhbar}(k)\). Therefore if we define
a mother wavelet to be the identity operator \(I\) the inverse
wavelet transform (cf.~\eqref{eq:inv-wavelet-transform}) will be
\begin{eqnarray}
  \label{eq:inverse-wavelet-convolution}
  \oper{M}_r {a} &=& \myhbar^n\int_{\Space{R}{2n}}
  a(x,y)\uir{b\myhbar}(\mathbf{s}((0,x,y)^{-1})) I\,dx\,dy\\
  &=& \myhbar^n \int_{\Space{R}{2n}}
  a(x,y)\uir{\myhbar}(0,x,y) \,dx\,dy\nonumber.
\end{eqnarray}
The inversion formula for \(\Space{H}{n}\) insures that 
\begin{prop}
  The composition \(\oper{M}_r\circ\oper{W}_r\) is the identity map on
  the representations \(\uir{\myhbar}(k)\) of convolution operators on
  \(\orbit{\myhbar}\). 
\end{prop}
\begin{example}
  \label{ex:wavelet-transform-operator}
  The wavelet transform \(\oper{W}_r\)~\eqref{eq:inversion-formula}
  applied to the quantum coordinate \(Q=d\uir{\myhbar}(X)\), momentum
  \(P=d\uir{\myhbar}(Y)\) (see~\eqref{eq:der-repr-h-bar}), and the
  energy function of the harmonic oscillator \((c_1Q^2+c_2P^2)/2\) produces
  the distributions on \(\Space{R}{2n}\):
  \begin{eqnarray*}
    Q  & \mapsto & \frac{1}{2\pi \rmi}\delta^{(1)}(x)\delta(y),\\
    P  & \mapsto & \frac{1}{2\pi \rmi}\delta(x)\delta^{(1)}(y),\\
    \frac{1}{2}\left(c_1 Q^2+c_2 P^2\right) & \mapsto &  
    -\frac{1}{8\pi^2} \left(c_1 \delta^{(2)}(x)\delta(y)
      +c_2 \delta(x)\delta^{(2)}(y)\right), \\
  \end{eqnarray*} where \(\delta^{(1)}\) and \(\delta^{(2)}\) are the
  first and second derivatives of the Dirac \emph{delta function} \(\delta\)
  respectively. The constants \(c_1\) and \(c_2\) have units
   \(M/T^2 \) and \(1/M\) correspondingly. We will use these
   distributions later in
  Example~\ref{ex:quantum-to-p-mechanics}.
\end{example}

\subsection{From Classical and Quantum Observables to $p$-Mechanics}
\label{sec:weyl-quantisation}

It is commonly accepted that we cannot deal with quantum mechanics
directly and thus classical dynamics serves as an unavoidable
intermediate step. A passage from classical observables to quantum
ones---known as a \emph{quantisation}---is the huge field with many
concurring approaches (geometric, deformation, Weyl, Berezin,
etc. quantisations) each having its own merits and demerits. Similarly
one has to construct \(p\)-mechanical observables starting from
classical or quantum ones by some procedure (should it be named
``\(p\)-mechanisation''?), which we about to describe now.

The transition from a \(p\)-mechanical observable to the classical one is
given by the formula~\eqref{eq:classical-observables}, which in a turn
is a realisation of the inverse wavelet
transform~\eqref{eq:inv-wavelet-transform}: 
\begin{equation}
  \label{eq:classical-observables-1}
  \uir{(q,p)}k=\hat{k}\left(0,{q},{p}\right)
  =c_\myhbar^{n+1}\int_{\Space{H}{n}}   k(s,x,y)e^{-{2\pi\rmi}(qx+py)} \,ds\,dx\,dy.
\end{equation} 

Similarly to a case of quantisation the classical image
\(\uir{(q,p)}k\)~\eqref{eq:classical-observables-1} contains only a
partial information about \(p\)-observable \(k\) unless we make
some additional assumptions.  Let us start from a classical observable
\(c(q,p)\) and construct the corresponding \(p\)-observable.
From the general consideration (see~\cite{Kisil98a} and
Section~\ref{sec:fock-type-spaces}) we can partially invert the
formula~\eqref{eq:classical-observables-1} by the wavelet
transform~\eqref{eq:wavelet-transform}:
\begin{equation}
  \label{eq:inv-classical-observables}
  \check{c}(x,y) = [\oper{W}_{0} c] (x,y)= \scalar{c v_{(0,0)}}{v_{(x,y)}}
  =c_\myhbar^n
  \int_{\Space{R}{2n}}   c(q,p)e^{{2\pi\rmi}(qx+py)} \,dq\,dp,
\end{equation} 
where \(v_{(x,y)}=\uir{(q,p)}v_{(0,0)}=e^{-2\pi\rmi(qx+py)}\).

However the function
\(\check{c}(x,y)\)~\eqref{eq:inv-classical-observables} is not defined
on the entire \(\Space{H}{n}\). The natural domain of
\(\check{c}(x,y)\) according to the construction of the reduced
wavelet transform~\cite{Kisil98a} is the homogeneous space
\(\Omega=G/Z\), where \(G=\Space{H}{n}\) and \(Z\) is its normal
subgroup of central elements \((s,0,0)\).  Let \(\mathbf{s}: \Omega
\rightarrow G\) be a Borel section in the principal bundle \(G
\rightarrow \Omega\), which is used in the construction of induced
representation, see~\cite[\S~13.1]{Kirillov76}. For the Heisenberg
group~\cite[Ex.~4.3]{Kisil98a} it can be simply defined as
\(\mathbf{s}: (x,y)\in\Omega\mapsto (0,x,y)\in\Space{H}{n} \). One can
naturally transfer functions from \(\Omega\) to the image
\(\mathbf{s}(\Omega)\) of the map \(\mathbf{s}\) in \(G\). However the
range \(\mathbf{s}(\Omega)\) of \(\mathbf{s}\) has oftenly
(particularly for \(\Space{H}{n}\)) a zero Haar measure in
\(G\). Probably two simplest possible ways out are:
\begin{enumerate}
\item To increase the ``weight'' of function \(\tilde{c}(s,x,y)\)
  vanishing outside of the range \(\mathbf{s}(\Omega)\) of
  \(\mathbf{s}\) by a suitable Dirac delta function on the subgroup
  \(Z\). For the Heisenberg group this can be done, for example, by
  the map:
  \begin{equation}
    \label{eq:p-mechanisation}
    \oper{E}: \check{c}(x,y)\mapsto \tilde{c}(s,x,y)=\delta(s)\check{c}(x,y), 
  \end{equation} where \(\check{c}(x,y)\) is given by the inverse
  wavelet (Fourier) transform~\eqref{eq:inv-classical-observables}. As
  we will see in Proposition~\ref{prop:Weyl-quantisation-decomposed}
  this is related to the Weyl quantisation and the Moyal brackets.
\item
  To extend the function \(\check{c}(x,y)\) to the entire group \(G\)
  by a tensor product with a suitable function on \(Z\), for example
  \(e^{-s^2}\):
  \begin{displaymath}
    \check{c}(x,y)\mapsto \tilde{c}(s,x,y)=e^{-s^2}\check{c}(x,y).
  \end{displaymath} In order to get the \emph{correspondence principle}
  between classical and quantum mechanics
  (cf. Example~\ref{ex:density-of-representations}) the function on \(Z\)
  has to satisfy some additional requirements. For \(\Space{H}{n}\) it
  should vanish for \(s\rightarrow\pm\infty\), which fulfils for both
  \(e^{-s^2}\) and \(\delta(s)\) from the previous item. In this way
  we get infinitely many essentially different quantisations with
  non-equivalent \emph{deformed} Moyal brackets between observables. 
\end{enumerate}
There are other more complicated possibilities not mentioned here,
which can be of some use if some additional information or
assumptions are used to extend functions from \(\Omega\) to
\(G\).  We will focus here only on the first ``minimalistic''
approach from the two listed above.

\begin{example}
  \label{ex:harmonic-oscillator-energy} The composition of the wavelet
  transform \(\oper{W}_0\)~\eqref{eq:inv-classical-observables}
  and the map \(\oper{E}\)~\eqref{eq:p-mechanisation} applied to the
  classical coordinate, momentum, and the energy function of a
  harmonic oscillator produces the distributions on \(\Space{H}{n}\):
  \begin{eqnarray}
    q  & \mapsto & \frac{1}{2\pi \rmi}\delta(s)\delta^{(1)}(x)\delta(y), \label{eq:p-mech-q}\\
    p  & \mapsto & \frac{1}{2\pi \rmi}\delta(s)\delta(x)\delta^{(1)}(y),\label{eq:p-mech-p}\\
    \frac{1}{2}\left(c_1 q^2+c_2 p^2\right) & \mapsto &  
    -\frac{1}{8\pi^2} \left(c_1 \delta(s)\delta^{(2)}(x)\delta(y)
      +c_2 \delta(s)\delta(x)\delta^{(2)}(y)\right), \label{eq:p-mech-q2+p2}
  \end{eqnarray} where \(\delta^{(1)}\), \(\delta^{(2)}\), \(c_1\),
  and \(c_2\) are defined in
  Example~\ref{ex:wavelet-transform-operator}.  We will use these
  distributions later in the Example~\ref{ex:harmonic-oscillator}.
\end{example}

\begin{figure}[tbp]
  \begin{center}
    \includegraphics{ubracket-unit.15}
  \end{center}
    \caption[Quantisation and wavelet transforms]{The relations
      between:\\
       \(\oper{Q}_\myhbar\)---the Weyl quantisation from classical
       mechanics to quantum;\\
       \(\oper{C}_{\myhbar\rightarrow 0}\)---the classical limit 
       \(\myhbar\rightarrow 0\) of       quantum mechanics;\\
       \(\uir{\myhbar}\) and \(\uir{(q,p)}\)---unitary representations
       of Heisenberg group \(\Space{H}{n}\); \\ 
       \(\oper{W}_r\) and \(\oper{W}_0\)---wavelet
      transforms defined in~\eqref{eq:inversion-trace} and~\eqref{eq:inv-classical-observables};\\ 
      \(\oper{E}\)---extension
       of      functions from \(\Omega=\Space{H}{n}/Z\) to
      the whole group \(\Space{H}{n}\).\\ 
      Note the relations \(\oper{Q}_\myhbar  =\uir{\myhbar}\circ\oper{E}\circ\oper{W}_0\) 
      and  \(C_{\myhbar\rightarrow  0} = \uir{(q,p)}\circ\oper{E}\circ\oper{W}_r\).}
    \label{fi:quant-class-p-observ}
\end{figure}  

If we apply the representation
\(\uir{\myhbar}\)~\eqref{eq:rho-extended-to-L1} to the function
\(\tilde{c}(s,x,y)\)~\eqref{eq:p-mechanisation} we will get the
operator on \(\FSpace{F}{2}(\orbit{\myhbar})\):
\begin{eqnarray}
  \oper{Q}_\myhbar (c)&=&c_\myhbar^{n+1} \int_{\Space{H}{n}}
  \tilde{c}(s,x,y)\uir{\myhbar}(s,x,y)\,ds\,dx\,dy \nonumber \\
  &=&c_\myhbar^{n+1} \int_{\Space{R}{2n}} \int_{\Space{R}{}}
  \delta(s)\check{c}(x,y)e^{s \cdot d\uir{\myhbar}(S) +x \cdot d\uir{\myhbar}(X)+
    y \cdot d\uir{\myhbar}(Y) }\,ds\,dx\,dy \nonumber \\
  &=&c_\myhbar^{n+1}  \int_{\Space{R}{}}\delta(s)e^{-2\pi\rmi s\myhbar}ds 
  \int_{\Space{R}{2n}}
  \check{c}(x,y)e^{x \cdot d\uir{\myhbar}(X)+
    y \cdot d\uir{\myhbar}(Y) }\,dx\,dy \nonumber  \\
  &=&c_\myhbar^{n}  \int_{\Space{R}{2n}}
  \check{c}(x,y)e^{x \cdot d\uir{\myhbar}(X)+
    y \cdot d\uir{\myhbar}(Y) }\,dx\,dy,   \label{eq:weyl-calculus}
\end{eqnarray} where the last expression is exactly the \emph{Weyl
quantisation} (the \emph{Weyl
correspondence}~\cite[\S~2.1]{Folland89}) if the Schr\"odinger
realisation with \(d\uir{\myhbar}(X)=q\) and
\(d\uir{\myhbar}(Y)=\rmi\myhbar \partial_q\) on
\(\FSpace{L}{2}(\Space{R}{n})\) is chosen for \(\uir{\myhbar}\). Thus
we demonstrate that

\begin{prop}
  \label{prop:Weyl-quantisation-decomposed} The \emph{Weyl
  quantisation} \(\oper{Q}_\myhbar\)~\eqref{eq:weyl-calculus} is the
  composition of the wavelet
  transform~\eqref{eq:inv-classical-observables}, the extension
  \(\oper{E}\)~\eqref{eq:p-mechanisation}, and the representation
  \(\uir{\myhbar}\)~\eqref{eq:stone-inf}:
  \begin{equation}
    \label{eq:Weyl-quantisation-decomposition}
    \oper{Q}_\myhbar =\uir{\myhbar}\circ\oper{E}\circ\oper{W}_0.
  \end{equation}
\end{prop}

The similar construction can be carried out if we have a
quantum observable \(A\) and wish to recover a related
\(p\)-mechanical object. The wavelet transform
\(\oper{W}_r\)~\eqref{eq:inversion-trace} 
maps \(A\) into the
function \(a(x,y)\) defined on \(\Omega\) and we again face the
problem of extension of \(a(x,y)\) to the entire group \(\Space{H}{n}\). If
it will be once more solved as in the classical case by the tensor product
with the delta function \(\delta(s)\) then we get the following
formula:
\begin{displaymath}
  A \mapsto a(s,x,y)=\oper{E}\circ\oper{W}_r (A)=\myhbar^{n}\delta(s)
  \int_{\Space{R}{2n}} \scalar[\FSpace{F}{2}(\orbit{\myhbar})]{A 
    v_{(x',y')}}{v_{(x,y)\cdot(x',y')}}\,dx'dy'.
\end{displaymath} We can apply to this function
\(a(s,x,y)\) the representation
\(\uir{(q,p)}\) and obtain a classical observables
\(\uir{(q,p)}(a)\). For a reasonable quantum observable \(A\) its
classical image \(\uir{(q,p)}\circ\oper{E}\circ\oper{W}_r (A)\) will
coincide with its classical limit \(C_{\myhbar\rightarrow
  0}A\):
\begin{equation}
  \label{eq:classical-limit-decomposition}
  C_{\myhbar\rightarrow  0} = \uir{(q,p)}\circ\oper{E}\circ\oper{W}_r,
\end{equation}
which is expressed here through integral transformations and does
not explicitly use any limit transition for \(\myhbar\rightarrow 0\).
The Figure~\ref{fi:quant-class-p-observ} illustrates various
transformations between quantum, classical, and
\(p\)-observables. Besides the mentioned
decompositions~\eqref{eq:Weyl-quantisation-decomposition}
and~\eqref{eq:classical-limit-decomposition} there are presentations of
identity maps on classical and quantum spaces correspondingly:
\begin{displaymath}
\oper{I}_c=\uir{(q,p)}\circ\oper{E}\circ\oper{W}_0, \qquad  
\oper{I}_\myhbar=\uir{\myhbar}\circ\oper{E}\circ\oper{W}_\myhbar.
\end{displaymath}
\begin{example}
  \label{ex:quantum-to-p-mechanics} The wavelet transform
  \(\oper{W}_r\) applied to the quantum coordinate \(Q\),
  momentum \(P\), and the energy function of a harmonic oscillator
  \((c_1Q^2+c_2P^2)/2\) was calculated in
  Example~\ref{ex:wavelet-transform-operator}. The composition with the
  above map \(\oper{E}\) yields the distributions:
  \begin{eqnarray*}
    Q  & \mapsto & \frac{1}{2\pi\rmi}\delta(s)\delta^{(1)}(x)\delta(y), \label{eq:p-mech-Q}\\
    P  & \mapsto & \frac{1}{2\pi\rmi}\delta(s)\delta^{(1)}(x)\delta(y) ,\label{eq:p-mech-P}\\
    \frac{1}{2}\left(c_1 Q^2+c_2 P^2\right) & \mapsto &  
    -\frac{1}{2\pi^2} \left(c_1 \delta(s)\delta^{(2)}(x)\delta(y)
      +c_2\delta(s)\delta(x)\delta^{(2)}(y)\right), \label{eq:p-mech-Q2+P2}
  \end{eqnarray*}
  which are exactly the same as in the
  Example~\ref{ex:harmonic-oscillator-energy}.
\end{example}

\section{$p$-Mechanics: Dynamics}
\label{sec:p-mechanics-dynamics}
We introduce the \(p\)-mechanical brackets, which suit to all essential
physical requirements and have a non-trivial classical representation
coinciding with the Poisson brackets. A consistent \(p\)-mechanical
dynamic equations is given in Subsection~\ref{sec:p-dynamic-equation}
and is analysed for the harmonic oscillator. Symplectic automorphisms of
the Heisenberg group produce symplectic symmetries of
\(p\)-mechanical, quantum, and classical dynamics in
Subsection~\ref{sec:quant-from-sypl}. 

\subsection{$p$-Mechanical Brackets on $\Space{H}{n}$}
\label{sec:p-mechanical-bracket}

Having observables as convolutions on \(\Space{H}{n}\) we need a
dynamic equation for their time evolution. To this end we seek a time
derivative generated by an observable associated with energy. 
\begin{rem}
  \label{re:defects-of-commutator}
  The first candidate is the derivation coming from
  commutator~\eqref{eq:commutator}. However the straight commutator
  has at least two failures:
  \begin{enumerate}
  \item It cannot produce any dynamics on
    \(\orbit{0}\)~\eqref{eq:orbit-0}, see
    Remark~\ref{re:triv-commutator}.
  \item It violates the Convention~\ref{conv:units-on-Hn} as indicated bellow.
  \end{enumerate} As well known the classical energy is measured in
  \(ML^2/T^2\) so does the \(p\)-mechanical energy \(E\). Consequently
  the commutator \([E,\cdot]\)~\eqref{eq:commutator} with the
  \(p\)-energy has units \(ML^2/T^2\) whereas the time derivative
  should be measured in \(1/T\), i.e. the mismatch is in the units of
  action \(ML^2/T\).
\end{rem}
Fortunately, there is a possibility to fix the both above defects of
the straight commutator at the same time.
Let us define a multiple \(\anti\) of a right inverse operator to the vector field
\(S\)~\eqref{eq:h-lie-algebra} on \(\Space{H}{n}\) by its actions on
exponents---characters of the centre \(Z\in\Space{H}{n}\): 
\begin{equation}
  \label{eq:def-anti}
 S\anti=4\pi^2 I, \qquad \textrm{ where }\quad
  \anti e^{ 2\pi\rmi \myhbar s}=\left\{ 
    \begin{array}{ll}
      \displaystyle \frac{2\pi}{\rmi\myhbar\strut} e^{2\pi\rmi\myhbar
      s}, & \textrm{if } \myhbar\neq 0,\\ 4\pi^2 s\strut, & \textrm{if
      } \myhbar=0.
    \end{array}
    \right. 
\end{equation} An alternative definition of \(\anti\) as the convolution
with a distribution is given in~\cite{Kisil00a}.

We can extend \(\anti\) by the linearity to the entire space
\(\FSpace{L}{1}(\Space{H}{n})\). As a multiplier of a right inverse to
\(S\) the operator \(\anti\) is measured in \(T/(ML^2)\)---exactly
that we need to correct the second of above mentioned defects of the
straight commutator.  Thus we introduce~\cite{Kisil00a} the modified
convolution operation \(\star\) on \(\FSpace{L}{1}(\Space{H}{n})\):
\begin{equation}
    \label{eq:u-star}
    k'\star k = (k'* k)\anti
\end{equation}
and the associated modified commutator (\(p\)-mechanical brackets):
\begin{equation}
  \label{eq:u-brackets}
    \ub{k'}{k}=[k',k]\anti=k'\star k-k\star k'.
\end{equation}
Obviously~\eqref{eq:u-brackets} is a bilinear antisymmetric form on
the convolution kernels. It was also demonstrated in~\cite{Kisil00a}
that \(p\)-mechanical brackets satisfy to the Leibniz and Jacoby
identities. They are all important for a consistent
dynamics~\cite{CaroSalcedo99} along with the dimensionality condition
given in the beginning of this Subsection. 

From~\eqref{eq:rho-extended-to-L1} one gets
\(\uir{\myhbar}(\anti k)=\frac{2\pi}{i\myhbar}\uir{\myhbar}(k)\) for
\(\myhbar\neq 0\). Consequently the modification
of the commutator for \(\myhbar\neq0\) is only slightly
different from the original one:
\begin{equation}
  \label{eq:rho-h-repres-of-p-bracket}
  \uir{\myhbar}\ub{k'}{k} =
  \frac{1}{\rmi\hbar}[\uir{\myhbar}(k'),\uir{\myhbar}(k)], \qquad
  \textrm{ where } \hbar=\frac{\myhbar}{2\pi}\neq0.
\end{equation} The integral representation of the modified commutator
kernel become (cf.~\eqref{eq:repres-commutator}):
\begin{equation}
  \fl
  \label{eq:repres-ubracket}
   \ub{k'}{k}\!\hat{_s}
    =   c_\myhbar^{n}\int_{\Space{R}{2n}}\! 
  \frac{4\pi}{\myhbar}\sin\left(\pi\myhbar  (xy'-yx')\right)
  \hat{k}'_s(\myhbar ,x',y')
  \hat{k}_s(\myhbar ,x-x',y-y') \,dx'dy', 
\end{equation}
where we can understand the expression under the integral  as
\begin{equation}
  \label{eq:convolution-kernel}
  \frac{4\pi}{\myhbar}\sin\left({\pi\myhbar}  (xy'-yx')\right)
  =4\pi^2\sum_{k=1}^\infty (-1)^{k+1}
  \left(\pi\myhbar\right)^{2(k-1)}
  \frac{   (xy'-yx')^{2k-1}}{(2k-1)!}  
\end{equation} This makes the operation~\eqref{eq:repres-ubracket} for
\(\myhbar=0\) significantly distinct from the vanishing
integral~\eqref{eq:repres-commutator}. Indeed it is natural to assign
the value \(4\pi^2(xy'-yx')\) to~\eqref{eq:convolution-kernel} for
\(\myhbar=0\). Then the integral in~\eqref{eq:repres-ubracket} becomes
the Poisson brackets for the Fourier transforms of \(k'\) and \(k\)
defined on \(\orbit{0}\)~\eqref{eq:orbit-0}:
\begin{equation}
  \label{eq:Poisson}
    \uir{(q,p)}\ub{k'}{k} = \frac{\partial \hat{k}'(0,q,p)}{\partial q}
    \frac{\partial \hat{k}(0,q,p)}{\partial p}
    -\frac{\partial \hat{k}'(0,q,p)}{\partial p} 
    \frac{\partial \hat{k}(0,q,p)}{\partial
      q}.
\end{equation} The same formula is obtained~\cite[Prop.~3.5]{Kisil00a}
if we directly calculate \(\uir{(q,p)}\ub{k'}{k}\) rather than resolve
the indeterminacy for \(\myhbar= 0\)
in~\eqref{eq:convolution-kernel}. This means the continuity of our
construction at \(\myhbar=0\) and represents the \emph{correspondence
principle} between quantum and classical mechanic.

We saw that the remedy of the second failure of commutator in
Remark~\ref{re:defects-of-commutator} (which was our duty according to
Convention~\ref{conv:units-on-Hn}) by the antiderivative~\eqref{eq:def-anti}
improves the first defect as well (which is a very pleasant and
surprising bonus). There are probably much simpler ways to fix the
dimensionality of commutator ``by hands''. However not all of them
obviously would produce the Poisson brackets on \(\orbit{0}\) as the
antiderivative~\eqref{eq:def-anti}.

We arrived to the following 
observation: \emph{Poisson brackets 
  and inverse of the Planck constant \(1/\myhbar\) have the same
  dimensionality because they are image of
the same object (anti-derivative~\eqref{eq:def-anti}) under different
representations \eqref{eq:stone-inf} and \eqref{eq:stone-one} of the
Heisenberg group.}

Note that functions \(X=\delta(s)\delta^{(1)}(x)\delta(y)\) and
\(Y=\delta(s)\delta(x)\delta^{(1)}(y)\) (see~\eqref{eq:p-mech-q}
and~\eqref{eq:p-mech-p}) on \(\Space{H}{n}\) are measured in units
\(L\) and \(ML^2/T\) (inverse to \(x\) and \(y\)) correspondingly
because they are respective derivatives of the dimensionless function
\(\delta(s)\delta(x)\delta(y)\). Then the \(p\)-mechanical brackets
\(\ub{X}{\cdot}\) and \(\ub{Y}{\cdot}\) with these functions have
dimensionality of \(T/(ML^2)\) and \(1/L\) correspondingly. Their
representation \(\uir{*} \ub{X}{\cdot}\) and \(\uir{*} \ub{Y}{\cdot}\)
(for both type of representations \(\uir{\myhbar}\) and \(\uir{(q,p)}\))
are measured by \(L\) and \(ML^2/T\) and are simple derivatives:
\begin{equation}
  \label{eq:shifts-on-orbits}
  \uir{*} \ub{X}{\cdot} = \frac{\partial }{\partial p}, \qquad  
  \uir{*} \ub{Y}{\cdot} = \frac{\partial }{\partial q} .
\end{equation} Thus \(\uir{*} \ub{X}{\cdot}\) and \(\uir{*}
\ub{Y}{\cdot}\) are generators of shits on both types of orbits
\(\orbit{\myhbar}\) and \(\orbit{0}\) independent from value of
\(\myhbar\).

\subsection{$p$-Mechanical Dynamic Equation}
\label{sec:p-dynamic-equation}

Since the modified commutator~\eqref{eq:u-brackets} with a
\(p\)-mechanical energy has the dimensionality \(1/T\)---the same as
the time derivative---we introduce the dynamic equation for an
observable \(f(s,x,y)\) on \(\Space{H}{n}\) based on that modified
commutator as follows
\begin{equation}
  \fl
  \label{eq:p-equation}
  \frac{d f}{d t}=\ub{f}{E} .
\end{equation}
\begin{rem}
  It is a general tendency to make a Poisson brackets or quantum
  commutator out of any two observables and say that they form a Lie
  algebra. However there is a physical meaning to do that if at least
  one of two observables is an energy, coordinate or momentum: in
  these cases the brackets produce the time
  derivative~\eqref{eq:p-equation} or corresponding shift
  generators~\eqref{eq:shifts-on-orbits}~\cite{HudsonPeck79a} of the
  other observable. 
\end{rem}
A simple consequence of the previous consideration is that the \(p\)-dynamic
equation~\eqref{eq:p-equation} is reduced 
\begin{enumerate}
\item by the representation \(\uir{\myhbar}\), \(\myhbar\neq0\)~\eqref{eq:stone-inf} on
        \(\FSpace{F}{2}(\orbit{\myhbar})\)~\eqref{eq:co-adjoint-orbits-inf}
        to Moyal's form of Heisenberg equation \cite[(8)]{Zachos02a}
        based on the formulae~\eqref{eq:rho-h-repres-of-p-bracket}
        and~\eqref{eq:repres-ubracket}:
        \begin{equation}
          \label{eq:moyal-equation}
          \frac{d \uir{\myhbar}(f)}{d t}
          =\frac{1}{i\hbar}[\uir{\myhbar}(f), H_\myhbar ], \qquad \textrm{
            where the operator } H_\myhbar =\uir{\myhbar}(E);
        \end{equation}
\item by the representations \(\uir{(q,p)}\)~\eqref{eq:stone-one} on
        \(\displaystyle\orbit{0}\)~\eqref{eq:orbit-0} to Poisson's 
        equation \cite[\S~39]{Arnold91} based on the formula~\eqref{eq:Poisson}:
        \begin{equation}
          \label{eq:Hamilton-equation}
          \frac{d \hat{f}}{d t}=\{\hat{f}, H\}
          \qquad \textrm{ where the function } H(q,p)=\uir{(q,p)}E=\hat{E}\left(0,{q},{p}\right).
        \end{equation}
\end{enumerate}
The same connections are true for the solutions of the three
equations~\eqref{eq:p-equation}--\eqref{eq:Hamilton-equation}.

\begin{figure}[tbp]
  \begin{center}
    \includegraphics{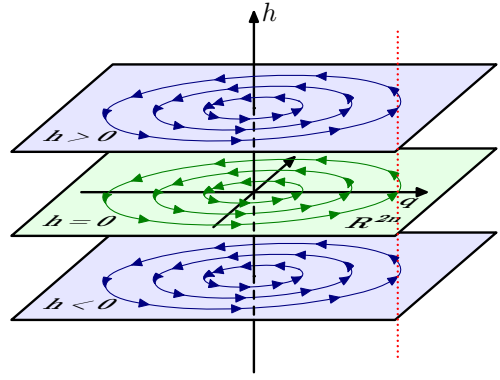}
  \end{center}
    \caption[Dynamics of the harmonic oscillator]{Dynamics of the
      harmonic oscillator in the adjoint space \(\algebra{h}_n^*\) is
      given by the identical linear symplectomorphisms of all orbits
      \(\orbit{\myhbar}\) and \(\orbit{0}\). The vertical dotted
      string is uniformly rotating in the ``horisontal'' plane around
      the \(h\)-axis without any dynamics along the
      ``vertical'' direction.}
  \label{fi:harmonic-scillator}
\end{figure}  

\begin{example}[harmonic oscillator, of course ;-]  
  \label{ex:harmonic-oscillator}\cite{Kisil00a}
  Let  the \(p\)-mechanical energy function of a
  harmonic oscillator be as obtained in
  Examples~\ref{ex:harmonic-oscillator-energy} and~\ref{ex:quantum-to-p-mechanics}: 
  \begin{equation}\label{eq:H-one}
    E(s,x,y)= -\frac{1}{8\pi^2} \left(c_1 \delta(s)\delta^{(2)}(x)\delta(y)
      +c_2 \delta(s)\delta(x)\delta^{(2)}(y)\right),
  \end{equation}
  Then the \(p\)-dynamic
  equation~\eqref{eq:p-equation} on \(\Space{H}{n}\) obeying the
  Convention~\ref{conv:units-on-Hn} is  
  \begin{equation}
    \label{eq:p-oscillator}
    \frac{d}{dt} f(t;s,x,y)= 
    \sum_{j=1}^n \left(c_2 x_j \frac{\partial}{\partial y_j} -
      c_1 y_j\frac{\partial}{\partial x_j}\right) f(t;s,x,y).
  \end{equation}
  Solutions to the above equation is well known to be 
  rotations in each of \((x_j,y_j)\) planes given by: 
  \begin{eqnarray}
    \label{eq:ho-evolution}
    f(t;s,x,y) & = & f_0\left(s, x\cos(\sqrt{c_1c_2}  t) -
      \sqrt{\frac{c_1}{c_2}}y\sin(\sqrt{c_1c_2}  t), \right. \\
    & & \qquad \left.
      \sqrt{\frac{c_2}{c_1}}x\sin(\sqrt{c_1c_2}  t) + y\cos (\sqrt{c_1c_2} t)\right).    
    \nonumber 
  \end{eqnarray} This expression respects the
  Convention~\ref{conv:units-on-Hn}. Since the dynamics on 
  \(\FSpace{L}{2}(\Space{H}{n})\) is given by a symplectic linear
  transformation of \(\Space{H}{n}\) its Fourier
  transform~\eqref{eq:fourier-transform} to
  \(\FSpace{L}{2}(\algebra{h}_n^*)\) is the adjoint symplectic
  linear transformation of orbits \(\orbit{\myhbar}\) and
  \(\orbit{0}\) in \(\algebra{h}_n^*\), see Figure~\ref{fi:harmonic-scillator}.

  The representation \(\uir{\myhbar}\)
  transforms the energy function \(E\)~\eqref{eq:H-one} into the operator
  \begin{equation}
    \label{eq:quantum-hamiltonian}
    H_\myhbar =-\frac{1}{8\pi^2}(c_1Q^2+c_2P^2),\quad
  \end{equation} where \( Q=d\uir{\myhbar}(X)\) and
  \(P=d\uir{\myhbar}(Y)\) are defined
  in~\eqref{eq:der-repr-h-bar}. The representation \(\uir{(q,p)}\)
  transforms \(E\) into the classical Hamiltonian
  \begin{equation}
    \label{eq:classical-hamilt}
    H (q,p)=\frac{c_1}{2} q^2+\frac{c_2}{2} p^2.
  \end{equation}
 
  The \(p\)-dynamic equation \eqref{eq:p-equation} in
  form \eqref{eq:p-oscillator} is transformed by the
  representations  \(\uir{\myhbar}\) into the Heisenberg equation 
  \begin{equation}
    \label{eq:heisenberg-equation}
    \frac{d}{dt} f(t;Q,P)=\frac{1}{i\hbar}[f,H_\myhbar ], \qquad
    \textrm{ where } \quad
    \frac{1}{i\hbar}[f,H_\myhbar ]=c_1p\frac{\partial
      f}{\partial q}-c_2q\frac{\partial f}{\partial p},
  \end{equation} defined by the operator
  \(H_\myhbar\)~\eqref{eq:quantum-hamiltonian}. The representation
  \(\uir{(q,p)}\) produces the Hamilton equation
  \begin{equation}
    \label{eq:hamilton-equation}
    \frac{d}{dt} f(t;q,p)=  c_1 p \frac{\partial f}{\partial q}-c_2 q \frac{\partial f}{\partial p}
  \end{equation} defined by the Hamiltonian
  \(H(q,p)\)~\eqref{eq:classical-hamilt}. Finally, to get the solution
  for equations~\eqref{eq:heisenberg-equation}
  and~\eqref{eq:hamilton-equation} it is enough to apply representations
  \(\uir{\myhbar}\) and \(\uir{(q,p)}\) to the
  solution~\eqref{eq:ho-evolution} of \(p\)-dynamic
  equation~\eqref{eq:p-oscillator}.
\end{example}

Summing up we can rephrase the title of~\cite{Zachos02a}:
\emph{quantum and classical mechanics live and work \textbf{together}
  on the Heisenberg group and are separated only in irreducible
  representations of   \(\Space{H}{n}\)}. 

\subsection{Symplectic Invariance from Automorphisms of  $\Space{H}{n}$}
\label{sec:quant-from-sypl}

\begin{figure}[tbp]
  \begin{center}
      \includegraphics{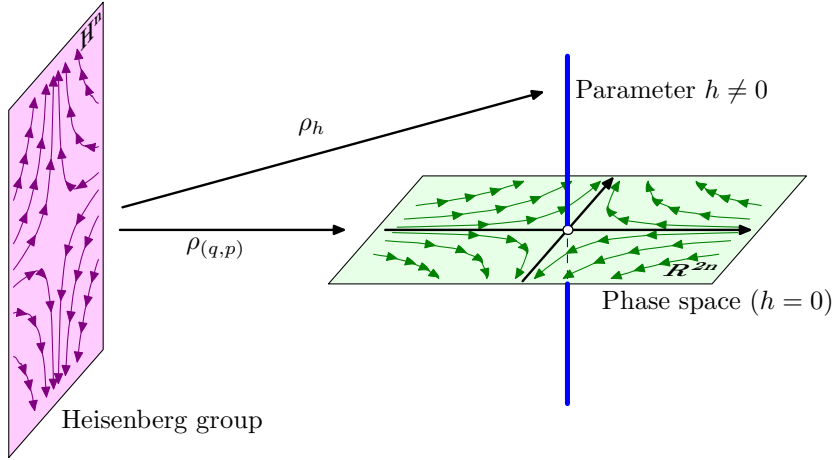}
  \end{center}
    \caption[Symplectic symmetries of quantum and classical mechanics]{
        \href{http://maths.leeds.ac.uk/~kisilv/nccg2.20.gif}{ 
          Automorphisms of
        \(\Space{H}{n}\) generated by the symplectic group \({Sp}(n)\) do
        not mix representations \(\uir{\myhbar}\) with different
        Planck constants \(\myhbar\) and act by the metaplectic
        representation inside each of them. In the contrast those
        automorphisms of \(\Space{H}{n}\) act transitively on the set
        of one-dimensional representations \(\uir{(q,p)}\) joining
        them into the tangent space of the classical phase space
        \(\Space{R}{2n}\)}.}
    \label{fig:p-mechanics}
\end{figure}

Let \(A: \Space{R}{2n} \rightarrow \Space{R}{2n}\) be a linear
\emph{symplectomorphism}~\cite[\S~41]{Arnold91}, \cite[\S~4.1]{Folland89},  i.e. a map defined
by \(2n\times 2n\) matrix:
\begin{displaymath}
  A: \twovect{x}{y} \mapsto \matr{a}{b}{c}{d}\twovect{x}{y}=\twovect{ax+by}{cx+dy}
\end{displaymath}
preserving the symplectic
form~\eqref{eq:symplectic-form}:
\begin{equation}
  \label{eq:symplectic-inavriance}
  \omega\left(A(x,y);A(x',y')\right)=\omega(x,y;x',y'). 
\end{equation}
All such transformations form the symplectic group \(Sp(n)\).
The Convention~\ref{conv:units-on-Hn} implies that  sub-blocks \(a\)
and \(d\) of \(A\) have to be dimensionless while \(b\) and \(c\) have
to be of reciprocal dimensions \(M/T\) and \(T/M\) respectively. 

It is follows from the identities~\eqref{eq:symplectic-inavriance}
and~\eqref{eq:H-n-group-law} that the the linear transformation
\(\alpha : \Space{H}{n} \rightarrow \Space{H}{n} \) such that
\(\alpha(s,x,y)=(s,A(x,y))\) is an automorphism of \(\Space{H}{n}\).
 Let us also denote by
\(\tilde{\alpha}=\tilde{\alpha}_A\)  a unitary transformation of
\(\FSpace{L}{2}(\Space{H}{n})\) in the form
\begin{displaymath}
  \tilde{\alpha}(f)(s,x,y)=\sqrt{\det a}f(s,A(x,y)),
\end{displaymath} which is well defined~\cite[\S~4.2]{Folland89} on
the double cover \(\widetilde{Sp}(n)\) of the group \(Sp(n)\).  The
correspondence \(A\mapsto \tilde{\alpha}_A\) is a linear unitary
representation of the symplectic group in
\(\FSpace{L}{2}(\Space{H}{n})\). One can also check the intertwining
property 
\begin{equation}
  \label{eq:alpha-intertwine-reg-repres}
  \lambda_{l(r)}(g)\circ \tilde{\alpha}=\tilde{\alpha}\circ
  \lambda_{l(r)}(\alpha(g)) 
\end{equation}
for the left (right) regular
representations~\eqref{eq:left-right-regular} of \(\Space{H}{n}\). 

Because \(\alpha\) is an automorphism
of \(\Space{H}{n}\) the map \(\alpha^*: k(g)\mapsto k(\alpha(g))\) is
an automorphism of the convolution algebra
\(\FSpace{L}{1}(\Space{H}{n})\) with the
multiplication~\(*\)~\eqref{eq:de-convolution}, i.e.
\({\alpha^*}(k_1)*{\alpha^*} 
(k_2)={\alpha^*}(k_1*k_2)\). Moreover \({\alpha^*}\) commutes
with the antiderivative~\(\anti\)~\eqref{eq:def-anti}, thus
\({\alpha}^*\) is an automorphism of
\(\FSpace{L}{1}(\Space{H}{n})\) with the modified
multiplication~\(\star\)~\eqref{eq:u-star} as well, i.e. \({\alpha^*}(k_1) 
\star{\alpha^*}(k_2)={\alpha^*}(k_1\star k_2)\).  By the linearity we
can extend the intertwining
property~\eqref{eq:alpha-intertwine-reg-repres} to the convolution
operator \(K\) as follows:
\begin{equation}
 \alpha^*K \circ\tilde{\alpha}=\tilde{\alpha}\circ K.
 \label{eq:intertwining}
\end{equation}

Since \(\alpha\) is an automorphism of \(\Space{H}{n}\) it fixes the
unit \(e\) of \(\Space{H}{n}\) and its differential
\(d\alpha:\algebra{h}^n\rightarrow \algebra{h}^n\) at \(e\) is given by the
same matrix as \(\alpha\) in the exponential coordinates. Obviously
\(d\alpha\) is an automorphism of the Lie algebra \(\algebra{h}^n\). 
By the duality between \(\algebra{h}^n\) and \(\algebra{h}^*_n\)
we obtain the adjoint map \(d\alpha^*:\algebra{h}^*_n\rightarrow
\algebra{h}^*_n\) defined by the expression
\begin{equation}
  \label{eq:d-alpha-star}
  d\alpha^*: (\myhbar,q,p) \mapsto (\myhbar,A^t(q,p)), 
\end{equation} where \(A^t\) is the transpose of \(A\).  Obviously
\(d\alpha^*\) preserves any orbit
\(\orbit{\myhbar}\)~\eqref{eq:co-adjoint-orbits-inf} and maps the
orbit \(\orbit{(q,p)}\)~\eqref{eq:co-adjoint-orbits-one} to
\(\orbit{A^t(q,p)}\). 

Identity~\eqref{eq:d-alpha-star} indicates that both representations
\(\uir{\myhbar}\) and \((\uir{\myhbar}\circ
\alpha)(s,x,y)=\uir{\myhbar}(s,A(x,y))\) for \(\myhbar\neq0\)
correspond to the same orbit \(\orbit{\myhbar}\). Thus they should be
equivalent,
i.e. there is an intertwining operator \(U_A:
\FSpace{F}{2}(\orbit{\myhbar}) \rightarrow
\FSpace{F}{2}(\orbit{\myhbar})\) such that \(U_A^{-1}
\uir{\myhbar}U_A=\uir{\myhbar}\circ \alpha\). Then the correspondence
\(\sigma: A \mapsto U_A\) is a linear unitary representation of the double
cover \(\widetilde{Sp}(n)\) of the symplectic group
called the \emph{metaplectic representation}~\cite[\S~4.2]{Folland89},
\cite{Hannabuss81a}. 

Thus we have
\begin{prop}
  \label{pr:symplectic}
  The \(p\)-mechanical brackets are invariant under the symplectic
  automorphisms of \(\Space{H}{n}\): \(\ub{\tilde{\alpha} k_1}{\tilde{\alpha} 
    k_2}= \tilde{\alpha}  \ub{k_1}{k_2}\). Consequently the dynamic
  equation~\eqref{eq:p-equation} has symplectic symmetries which are
  reduced 
\begin{enumerate}
\item by \(\uir{\myhbar}\), \(\myhbar\neq0\) on
  \(\orbit{\myhbar}\)~\eqref{eq:co-adjoint-orbits-inf} to the
  metaplectic representation in quantum mechanics;
\item by \(\uir{(q,p)}\) on
  \(\displaystyle\orbit{0}\)~\eqref{eq:orbit-0} to the symplectic 
  symmetries of classical mechanics~\textup{\cite[\S~38]{Arnold91}}.
\end{enumerate}
\end{prop}


Combining intertwining properties of all three
components~\eqref{eq:Weyl-quantisation-decomposition} in the Weyl
quantisation we get
\begin{cor}
  The Weyl quantisation \(\oper{Q}_\myhbar\)~\eqref{eq:weyl-calculus}
  is the \emph{intertwining operator} between classical and
  metaplectic representations.
\end{cor}

\section{Conclusions}
\label{sec:conclusions}

\subsection{Discussion}
\label{sec:discussion}

Our intention is to demonstrate that 
the complete
representation theory of the Heisenberg group \(\Space{H}{n}\), which
includes one-dimensional commutative representations, is a sufficient
language for \emph{both} classical and quantum theory.


It is natural to describe the complete set of unitary irreducible
representations by the orbit method of Kirillov. The analysis carried out
in Subsection~\ref{sec:struct-topol-unit} and illustrated in
Figure~\ref{fi:unitary-dual} shows that the position of
one-dimensional representations \(\uir{(q,p)}\) within the unitary
dual of \(\Space{H}{n}\) relates them to classical mechanics. Various
connections of infinite dimensional representations \(\uir{\myhbar}\)
of \(\Space{H}{n}\) to quantum mechanics has been known for a long time.

Convolutions operators on \(\Space{H}{n}\) is a natural class to
be associated with physical observables. They are reduced by infinite
dimensional representations \(\uir{\myhbar}\) to the
pseudodifferential operators, which are observables in the Weyl
quantisation. The one-dimensional representations \(\uir{(q,p)}\) map
convolutions into classical observables---functions on the phase
space. The wavelets technique allows us to transform these three types of
observables into each other, which is illustrated by
Figure~\ref{fi:quant-class-p-observ}.

A nontrivial dynamics in the phase space---the space of
one-dimensional representations of \(\Space{H}{n}\)---could be
obtained from the commutator on \(\Space{H}{n}\) with a help of the
anti-derivative operator \(\anti\)~\eqref{eq:def-anti}. The
\(p\)-mechanical dynamic equation~\eqref{eq:p-equation} based on the
operator \(\anti\) possesses all desirable properties for description of
a physical time evolution and its solution gives both classic and
quantum dynamics. See Figure~\ref{fi:harmonic-scillator} for the
familiar dynamics of the harmonic oscillator.

Finally, the symplectic automorphisms of the Heisenberg group preserve
the dynamic equation~\eqref{eq:p-equation} and all its solutions. In
representations of the Heisenberg group this reduces to the symplectic
invariance of classical mechanics and the metaplectic invariance of the
quantum description. Moreover the symplectic transformations act
transitively on the set \(\orbit{0}\)~\eqref{eq:orbit-0} of
one-dimensional representations supporting its \(p\)-mechanical
interpretation as the classical phase space, see
  Figure~\ref{fig:p-mechanics}. 

\subsection{Further Developments}
\label{sec:furth-devel}
The present paper deals only with elementary aspects of
\(p\)-mechanics. Notion of physical states in \(p\)-mechanics is
considered in~\cite{Brodlie03a,BrodlieKisil03a}, where its usefulness
for a forced oscillator is demonstrated. Paper~ \cite{BrodlieKisil03a}
discusses also connection of \(p\)-mechanics and \emph{contextual
  interpretation}~\cite{Khrennikov02a}. 
Our study is a part of the
Erlangen-type approach~\cite{Kisil97a,Kisil02c} in non-commutative
geometry.  It could be extended in several directions:

\subsubsection{Quantum-Classical Interaction}
\label{sec:test}
  The long standing discussion~\cite{CaroSalcedo99,Prezhdo-Kisil97}
  about quantum-classical interaction can be treated as
  follows. Let \(\Space{B}{}\) be a nilpotent step two Heisenberg-like
  group of elements \((s_1,s_2;x_1,y_1;x_2, y_2)\) with the only
  non-trivial commutators in the Lie algebra
  (cf.~\eqref{eq:heisenberg-comm}) as follows:
  \begin{displaymath}
    [X_i,Y_j]=\delta_{ij} S_i.
  \end{displaymath} Thus \(\Space{B}{}\) has the two dimensional
  centre \((s_1,s_2,0,0,0,0)\) and the adjoint space of characters of
  \(\Space{B}{}\) is also two dimensional.  We can regard it as being
  spanned by two different Planck constants \(\myhbar_1\) and
  \(\myhbar_2\). There is a possibility to study the case
  \(\myhbar_1\neq 0\) and \(\myhbar_2=0\), which correspond to a
  quantum behaviour of coordinates \((x_1,y_1)\) and a classical
  dynamics in \((x_2,y_2)\). This study was initiated
  in~\cite{Prezhdo-Kisil97} but oversaw some homological aspects of
  the construction and is not satisfactory completed yet.

\subsubsection{Quantum Field Theory}
\label{sec:quantum-field-theory}

  Mathematical formalism of quantum mechanics uses complex numbers in
  order to provide unitary infinite dimensional representations of the
  Heisenberg group \(\Space{H}{n}\).  In a similar way the
  De~Donder--Weyl formalism for classical field
  theories~\cite{Kanatchikov01b} requires \emph{Clifford
  numbers}~\cite{Hestenes99} for their quantisation.  It was recently
  realised~\cite{CnopsKisil97a} that the appearance of Clifford
  algebras is induced by the Galilean group---a nilpotent step two Lie
  group with multidimensional centre. In the one-dimensional case an
  element of the Galilean group is
  \((s_1,\ldots,s_n,x,y_1,\ldots,y_n)\) with corresponding the Lie algebra
  described by the non-vanishing commutators:
  \begin{displaymath}
    [X, Y_j]=S_j, \qquad j=1,2,\ldots,n.
  \end{displaymath} This corresponds to several momenta \(y_1\),
  \(y_2\), \ldots, \(y_n\) adjoint to a single field coordinate
  \(x\)~\cite{Kanatchikov01b}.  For field theories it is
  worth~\cite{Kisil03a} to consider Clifford valued representations
  induced by Clifford valued ``characters'' \(e^{2\pi(e_1\myhbar_1
  s_1+\cdots+e_n\myhbar_n s_n)}\) of the centre, where \(e_1\), \ldots
  \(e_n\) are imaginary units spanning the Clifford algebra. The
  associated Fock spaces were described in~\cite{CnopsKisil97a}.
  In~\cite{Kisil03a} we quantise the De~Donder--Weyl field equations
  (similarly to our consideration in
  Subsection~\ref{sec:p-mechanical-bracket}) with the help of composed
  antiderivative operator \(\anti=\sum_1^n e_i\anti_i\), where
  \(S_i\anti_i=4\pi^2I\).  There are important mathematical and
  physical questions related to the construction, notably the r\^ole
  of the Dirac operator~\cite{Getzler86a}, which deserve further
  careful considerations.

\subsubsection{String Theory}
\label{sec:string-theory}

   There is a possibility to use \(p\)-mechanical picture for a
   string-like theory. Indeed the \(p\)-dynamics of a harmonic
   oscillator as presented in the Example~\ref{ex:harmonic-oscillator}
   and Figure~\ref{fi:harmonic-scillator} consists of the uniform
   rotation of lines around the \(\myhbar\)-axis---one can say
   \emph{strings}---with the same \((q,p)\) coordinates but different
   values of the Planck constant\(\myhbar\).
   
   In case of a more general energy, which is still however given by a
   \emph{convolution} on \(\Space{H}{n}\), the dynamics can be more
   complicated. For example, it may not correspond to a point
   transformation of the adjoint space
   \(\algebra{h}_n^*\). Alternatively a generic point
   transformation may transform a straight line
   \((\myhbar,q_0,p_0)\) with fixed \((q_0,p_0)\in\Space{R}{2n}\) and
   variable \(\myhbar\) into a generic curve transversal to all
   \((q,p)\)-planes. However all spaces
   \(\FSpace{F}{2}(\orbit{\myhbar})\) are invariant under any
   \(p\)-dynamics generated by a convolution on \(\Space{H}{n}\).

   However if an energy is given by an arbitrary operator on
  \(\FSpace{L}{2}(\Space{H}{n})\)~\cite{Dynin75,Kisil96e} then spaces
  \(\FSpace{F}{2}(\orbit{\myhbar})\) for different \(\myhbar\) are no
  longer invariant during the evolution and could be mixed together. This
  opens a possibility of longitudinal dynamics of strings along the
  \(\myhbar\)-axis as well. It may seem strange to have a dynamics along
  \(\myhbar\) which is a \emph{constant}, not a variable. However
  there is a duality~\cite{Witten96} between the ``Planck constant''
  \(\myhbar\) and the ``tension of string'' \(\alpha'\). Dualities and
  symmetries between \(\myhbar\) and \(\alpha'\) can be reflected in
  a dynamics which mixes spaces \(\FSpace{F}{2}(\orbit{\myhbar})\)
  with different~\(\myhbar\). 


\section*{Acknowledgements}
\label{sec:acknowledgements}

I am grateful to A.~Brodlie for many useful discussions and
comments. Anonymous referees made numerous critical remarks and
suggestions, which resulted into the paper's improvements. 

\bibliographystyle{amsplain}
\bibliography{abbrevmr,akisil,analyse,aphysics}

\newcommand{\noopsort}[1]{} \newcommand{\printfirst}[2]{#1}
  \newcommand{\singleletter}[1]{#1} \newcommand{\switchargs}[2]{#2#1}
  \newcommand{\irm}{\textup{I}} \newcommand{\iirm}{\textup{II}}
  \newcommand{\vrm}{\textup{V}}
  \providecommand{\cprime}{'}\providecommand{\arXiv}[1]{\eprint{http://arXiv.o%
rg/abs/#1}{arXiv:#1}}
\providecommand{\bysame}{\leavevmode\hbox to3em{\hrulefill}\thinspace}
\providecommand{\MR}{\relax\ifhmode\unskip\space\fi MR }
\providecommand{\MRhref}[2]{%
  \href{http://www.ams.org/mathscinet-getitem?mr=#1}{#2}
}
\providecommand{\href}[2]{#2}
\begin{thebibliography}{10}

\bibitem{AliAntGaz00}
Syed~Twareque Ali, Jean-Pierre Antoine, and Jean-Pierre Gazeau, \emph{Coherent
  states, wavelets and their generalizations}, Graduate Texts in Contemporary
  Physics, Springer-Verlag, New York, 2000. \MR{2002m:81092}

\bibitem{Arnold91}
V.~I. Arnol{\cprime}d, \emph{Mathematical methods of classical mechanics},
  Graduate Texts in Mathematics, vol.~60, Springer-Verlag, New York, 1991,
  Translated from the 1974 Russian original by K. Vogtmann and A. Weinstein,
  Corrected reprint of the second (1989) edition. \MR{96c:70001}

\bibitem{Berezin72}
F.~A. Berezin, \emph{Covariant and contravariant symbols of operators}, Izv.
  Akad. Nauk SSSR Ser. Mat. \textbf{36} (1972), 1134--1167, Reprinted
  in~\cite[pp.~228--261]{Berezin86}. \MR{50 \#2996}

\bibitem{Berezin75b}
\bysame, \emph{General concept of quantization}, Comm. Math. Phys. \textbf{40}
  (1975), 153--174. \MR{53 \#15186}

\bibitem{Berezin86}
\bysame, \emph{Metod vtorichnogo kvantovaniya}, second ed., ``Nauka'', Moscow,
  1986, Edited and with a preface by M. K. Polivanov. \MR{89c:81001}

\bibitem{Brodlie03a}
Alastair Brodlie, \emph{Classical and quantum coherent states}, Internat. J.
  Theoret. Phys. (2003), 20~p., \arXiv{quant-ph/0303142} (To appear).

\bibitem{BrodlieKisil03a}
Alastair Brodlie and Vladimir~V. Kisil, \emph{Observables and sates in
  \(p\)-mechanics}, to appear, Advances in Mathematics Research, Nova Sci.,
  2003, \arXiv{quant-ph/0304023}.

\bibitem{CaroSalcedo99}
J.~Caro and L.~L. Salcedo, \emph{Impediments to mixing classical and quantum
  dynamics}, Phys. Rev. \textbf{A60} (1999), 842--852,
  \arXiv{quant-ph/9812046}.

\bibitem{CnopsKisil97a}
Jan Cnops and Vladimir~V. Kisil, \emph{Monogenic functions and representations
  of nilpotent {Lie} groups in quantum mechanics}, Mathematical Methods in the
  Applied Sciences \textbf{22} (1998), no.~4, 353--373, \arXiv{math/9806150}.
  \MR{2000b:81044}. \Zbl{1005.22003}.

\bibitem{Dynin75}
A.~S. Dynin, \emph{Pseudodifferential operators on the {H}eisenberg group},
  Dokl. Akad. Nauk SSSR \textbf{225} (1975), no.~6, 1245--1248. \MR{54 \#11410}

\bibitem{Folland89}
Gerald~B. Folland, \emph{Harmonic analysis in phase space}, Annals of
  Mathematics Studies, vol. 122, Princeton University Press, Princeton, NJ,
  1989. \MR{92k:22017}

\bibitem{Getzler86a}
Ezra Getzler, \emph{A short proof of the local {A}tiyah-{S}inger index
  theorem}, Topology \textbf{25} (1986), no.~1, 111--117. \MR{87h:58207}

\bibitem{Hannabuss81a}
K.~C. Hannabuss, \emph{Characters and contact transformations}, Math. Proc.
  Cambridge Philos. Soc. \textbf{90} (1981), no.~3, 465--476. \MR{83e:81035}

\bibitem{Hestenes99}
David Hestenes, \emph{New foundations for classical mechanics}, second ed.,
  Kluwer Academic Publishers Group, Dordrecht, 1999. \MR{2001h:70004}

\bibitem{Howe80b}
Roger Howe, \emph{Quantum mechanics and partial differential equations}, J.
  Funct. Anal. \textbf{38} (1980), no.~2, 188--254. \MR{83b:35166}

\bibitem{HudsonPeck79a}
R.~L. Hudson and S.~N. Peck, \emph{Canonical {F}ourier transforms}, J. Math.
  Phys. \textbf{20} (1979), no.~1, 114--119. \MR{81j:46107}

\bibitem{Kanatchikov01b}
Igor~V. Kanatchikov, \emph{Precanonical quantum gravity: quantization without
  the space-time decomposition}, Internat. J. Theoret. Phys. \textbf{40}
  (2001), no.~6, 1121--1149, \arXiv{gr-qc/0012074}. \MR{2002m:83038}

\bibitem{Khrennikov02a}
Andrei Khrennikov, \emph{V\"axj\"o interpretation of quantum mechanics},
  (2002), \arXiv{quant-ph/0202107}.

\bibitem{Kirillov76}
A.~A. Kirillov, \emph{Elements of the theory of representations},
  Springer-Verlag, Berlin, 1976, Translated from the Russian by Edwin Hewitt,
  Grundlehren der Mathematischen Wissenschaften, Band 220. \MR{54 \#447}

\bibitem{Kirillov94a}
\bysame, \emph{Introduction to the theory of representations and noncommutative
  harmonic analysis [{\MR{90a:22005}}]}, Representation Theory and
  Noncommutative Harmonic Analysis, I, Springer, Berlin, 1994, \MR{1311488}.,
  pp.~1--156, 227--234. \MR{1 311 488}

\bibitem{Kirillov99}
\bysame, \emph{Merits and demerits of the orbit method}, Bull. Amer. Math. Soc.
  (N.S.) \textbf{36} (1999), no.~4, 433--488. \MR{2000h:22001}

\bibitem{Kisil96e}
Vladimir~V. Kisil, \emph{Local algebras of two-sided convolutions on the
  {H}eisenberg group}, Mat. Zametki \textbf{59} (1996), no.~3, 370--381, 479,
  \MR{97h:22006}.

\bibitem{Kisil96a}
\bysame, \emph{Plain mechanics: Classical and quantum}, J. Natur. Geom.
  \textbf{9} (1996), no.~1, 1--14, \MR{96m:81112}. \arXiv{funct-an/9405002}.

\bibitem{Kisil97a}
\bysame, \emph{Two approaches to non-commutative geometry}, Complex Methods for
  Partial Differential Equations (H.~Begehr, O.~Celebi, and W.~Tutschke, eds.),
  Kluwer Academic Publishers, Netherlands, 1999, \arXiv{funct-an/9703001},
  \MR{2001a:01002}, pp.~219--248.

\bibitem{Kisil98a}
\bysame, \emph{Wavelets in {Banach} spaces}, Acta Appl. Math. \textbf{59}
  (1999), no.~1, 79--109, \arXiv{math/9807141}. \MR{2001c:43013}.

\bibitem{Kisil02c}
\bysame, \emph{Meeting {Descartes} and {Klein} somewhere in a noncommutative
  space}, Highlights of Mathematical Physics (A.~Fokas, J.~Halliwell,
  T.~Kibble, and B.~Zegarlinski, eds.), AMS, 2002, \arXiv{math-ph/0112059},
  pp.~165--189.

\bibitem{Kisil00a}
\bysame, \emph{Quantum and classical brackets}, Internat. J. Theoret. Phys.
  \textbf{41} (2002), no.~1, 63--77, \arXiv{math-ph/0007030}. \MR{2003b:81105}

\bibitem{Kisil03a}
\bysame, \emph{$p$-{Mechanics} and {De}~{Donder}--{Weyl} theory}, The Fifth
  International Conference ``Symmetry in Nonlinear Mathematical Physics'',
  Inst. of Math., NAS of Ukraine, 2003, \arXiv{quant-ph/0306101} (To appear).

\bibitem{Klauder94b}
John~R. Klauder, \emph{Coherent states and coordinate-free quantization},
  Internat. J. Theoret. Phys. \textbf{33} (1994), no.~3, 509--522.
  \MR{95a:81105}

\bibitem{Perelomov86}
A.~Perelomov, \emph{Generalized coherent states and their applications}, Texts
  and Monographs in Physics, Springer-Verlag, Berlin, 1986. \MR{87m:22035}

\bibitem{Prezhdo-Kisil97}
Oleg~V. Prezhdo and Vladimir~V. Kisil, \emph{Mixing quantum and classical
  mechanics}, Phys. Rev. A (3) \textbf{56} (1997), no.~1, 162--175,
  \MR{99j:81010}. \arXiv{quant-ph/9610016}.

\bibitem{Shubin87}
M.~A. Shubin, \emph{Pseudodifferential operators and spectral theory}, second
  ed., Springer-Verlag, Berlin, 2001, Translated from the 1978 Russian original
  by Stig I. Andersson. \MR{2002d:47073}

\bibitem{MTaylor86}
Michael~E. Taylor, \emph{Noncommutative harmonic analysis}, Mathematical
  Surveys and Monographs, vol.~22, American Mathematical Society, Providence,
  RI, 1986. \MR{88a:22021}

\bibitem{Witten96}
Edward Witten, \emph{Reflections on the fate of spacetime}, Phys. Today
  \textbf{49} (1996), no.~4, 24--30. \MR{97i:81003}

\bibitem{Woodhouse80}
N.~M.~J. Woodhouse, \emph{Geometric quantization}, second ed., Oxford
  Mathematical Monographs, The Clarendon Press Oxford University Press, New
  York, 1992, Oxford Science Publications. \MR{94a:58082}

\bibitem{Zachos02a}
Cosmas Zachos, \emph{Deformation quantization: quantum mechanics lives and
  works in phase-space}, Internat. J. Modern Phys. A \textbf{17} (2002), no.~3,
  297--316, \arXiv{hep-th/0110114}. \MR{1 888 937}

\end{thebibliography}

\end{document}